\documentclass[11pt,fleqn]{article}
\usepackage{epsfig}
\usepackage{rotating}
\usepackage{amsmath}
\usepackage{psfrag}
\allowdisplaybreaks[1]
 
\usepackage{subfigure}
\renewcommand{\thesubfigure}{(\arabic{subfigure})}

\newcommand{\nnl}{\nonumber\\}
\newcommand{\kk}[1]{_{(#1)}}  
\newcommand{\kkx}[2]{_{#1(#2)}}  

\newcommand{\mcA}{\mathcal{A}}
\newcommand{\mcD}{\mathcal{D}}

\newcommand{\mcG}{\mathcal{G}}
\newcommand{\mcL}{\mathcal{L}}
\newcommand{\mcQ}{\mathcal{Q}}
\newcommand{\mcU}{\mathcal{U}}
\newcommand{\fslash}[1]{#1\!\!\!/}  
\newcommand{\fslashl}[1]{#1\!\!\!\!/}  
\newcommand{\nor}{\frac{n}{R}} 
\newcommand{\norsq}{\frac{n^2}{R^2}} 
\newcommand{\GeV}{~\text{GeV}}
\newcommand{\dpkt}{\;:\quad}  
\newcommand{\sw}{s_w}

\newcommand{\RE}{{\rm Re}}
\newcommand{\IM}{{\rm Im}}

\newcommand{\vtd}{|V_{td}|}

\newcommand{\vts}{|V_{ts}|}

\def\R1{\varepsilon_1}
\def\E8{\varepsilon_8}

\def\epe{\varepsilon'/\varepsilon}
\def\as{\alpha_s}

\newcommand{\nn}{\nonumber}
\newcommand{\mt}{m_{\rm t}}

\newcommand{\mc}{m_{\rm c}}
\newcommand{\ms}{m_{\rm s}}
\newcommand{\md}{m_{\rm d}}
\newcommand{\mb}{m_{\rm b}}

\newcommand{\mw}{M_{\rm W}}
\newcommand{\mz}{M_{\rm Z}}

\newcommand{\gev}{\, {\rm GeV}}
\newcommand{\mev}{\, {\rm MeV}}
\newcommand{\bsi}{B_6^{(1/2)}}
\newcommand{\bei}{B_8^{(3/2)}}
\newcommand{\Lms}{\Lambda_{\overline{\rm MS}}}

\newcommand{\newsection}[1]{\section{#1}\setcounter{equation}{0}}
\newcommand{\bea}{\begin{eqnarray}}
\newcommand{\eea}{\end{eqnarray}}
\newcommand{\bd}{\begin{displaymath}}
\newcommand{\ed}{\end{displaymath}}

\newcommand{\be}{\begin{equation}}
\newcommand{\ee}{\end{equation}}
\newcommand{\bi}{\begin{itemize}}
\newcommand{\ei}{\end{itemize}}
\newcommand{\ord}{{\cal O}}

\newcommand{\Ctilde}{\tilde{C}}

\newcommand{\kpn}{K^+\rightarrow\pi^+\nu\bar\nu}
\newcommand{\klpn}{K_{\rm L}\rightarrow\pi^0\nu\bar\nu}

\newcommand{\klm}{K_{\rm L} \to \mu^+\mu^-}

\newcommand{\kpe}{K_{\rm L} \to \pi^0 e^+ e^-}

\renewcommand{\baselinestretch}{1.3}

\begin{document}
\thispagestyle{empty}
\phantom{xxx}
\vskip0.3truecm
\begin{flushright}
 TUM-HEP-512/03 \\
 MPI-PhT/2003-26 
\end{flushright}
\vskip0.4truecm

\begin{center}
 {\LARGE\bf The Impact of Universal Extra Dimensions on\\
 \boldmath{$B\to X_s\gamma$}, \boldmath{$B\to X_s~ {\rm gluon}$}, 
\boldmath{$B\to X_s\mu^+\mu^-$},\\ 
\boldmath{$K_L\to \pi^0e^+e^-$} and \boldmath{$\epe$}\\ }
\end{center}

\vskip1truecm
\centerline{\Large\bf Andrzej J. Buras${}^a$, 
Anton Poschenrieder${}^{a}$} 
\centerline{\Large\bf Michael Spranger${}^{a,b}$ and Andreas Weiler${}^a$} 
\bigskip
\centerline{\sl ${}^a$ Physik Department, Technische Universit{\"a}t M{\"u}nchen } 
\centerline{\sl D-85748 Garching, Germany}
\vspace{2ex}
\centerline{\sl ${}^b$ Max-Planck-Institut f{\"u}r Physik - Werner-Heisenberg-Institut}
\centerline{\sl D-80805 M{\"u}nchen, Germany}

\vskip1truecm
\centerline{\bf Abstract}
We calculate the contributions of the Kaluza-Klein (KK) modes  to 
the $\gamma$--penguins, gluon--penguins, $\gamma$--magnetic penguins and 
chromomagnetic penguins 
in the Appelquist, Cheng and Dobrescu (ACD) model with one universal extra 
dimension. 
Together with our previous calculation of $Z^0$ penguin diagrams 
\cite{BSW02} this allows to study the impact of the KK modes on the decays 
$B\to X_s\gamma$, $B\to X_s~{\rm gluon}$, $B\to X_s\mu^+\mu^-$ and 
$K_L\to \pi^0e^+e^-$ and on the CP-violating ratio $\epe$.
For the compactification 
scale $1/R= 300\gev$ the perturbative part of the branching ratio for 
$B\to X_s\mu^+\mu^-$ is
enhanced by $12\%$ while the zero in the $A_{\rm FB}$ asymmetry is shifted 
from $\hat s_0=0.162$ to $\hat s_0=0.142$. The sizable suppressions of 
$Br(B\to X_s\gamma)~(20\%)$ and $Br(B\to X_s~{\rm gluon})~(40\%)$ could 
have interesting phenomenological implications on the lower bound on 
$1/R$ provided the experimental 
and theoretical uncertainties will be decreased.
Similar comments apply to $\epe$ that is suppressed relative to the 
Standard Model expectations with the size of the suppression depending 
sensitively on the hadronic matrix elements. The impact on 
$K_L\to \pi^0e^+e^-$ is below $10\%$. We point out a correlation between
the zero $\hat s_0$ in  the $A_{\rm FB}$ asymmetry and 
$Br(B\to X_s\gamma)$ that should be valid in most models with minimal 
flavour violation.


\newpage

\section{Introduction}
\setcounter{equation}{0}
In a recent paper \cite{BSW02} we have calculated
the contributions of the Kaluza-Klein (KK) modes
to the $K_L-K_S$ mass difference $\Delta M_K$, the parameter $\varepsilon_K$, 
the $B^0_{d,s}-\bar B^0_{d,s}$ mixing mass differences $\Delta M_{d,s}$ and 
rare decays  
$\kpn$, $\klpn$, $\klm$, $B\to X_{s,d}\nu\bar\nu$ and $B_{s,d}\to\mu^+\mu^-$ 
in the Appelquist, Cheng and Dobrescu (ACD) model \cite{appelquist:01}
with one universal extra dimension.
In this model 
the only additional free parameter relative to the Standard Model (SM) 
is the compactification
scale $1/R$. Thus all the masses of the KK particles and their interactions 
among themselves and with the SM particles are described in terms of $1/R$ 
and the parameters of the SM. This economy in new parameters should be 
contrasted with supersymmetric theories and models with an extended Higgs 
sector.

A very important property of the ACD model is
the conservation of KK parity that implies the absence of tree level 
KK contributions to low energy processes taking place at scales $\mu\ll 1/R$.
In this context the  flavour changing neutral current (FCNC) processes 
considered in \cite{BSW02}  are of particular 
interest as they appear in the SM first at one-loop and are 
strongly suppressed. Consequently  the one-loop contributions from 
the KK modes to them could in principle be important.
Indeed our analysis in \cite{BSW02} has shown that for $1/R \le 400\gev$
the KK effects in some of the processes in question could be significant.
This is in particular the case of the decays governed by the 
CKM element $\vts$, like $B\to X_{s}\nu\bar\nu$ and
$B_{s}\to\mu^+\mu^-$, in which the enhancement through the KK modes in
$Z^0$ penguin diagrams is not softened by the suppression of the relevant 
CKM parameters in contrast to the processes governed by $\vtd$.

In the present paper we extend the analysis of \cite{BSW02} by calculating 
the decays that receive contributions from
the $\gamma$ penguins, gluon penguins, $\gamma$--magnetic penguins and 
chromomagnetic penguins. In particular we analyze in detail the impact 
of the KK contributions on the decays
$B\to X_s\gamma$, $B\to X_s~ {\rm gluon}$, $B\to X_s\mu^+\mu^-$ and 
$K_L\to \pi^0e^+e^-$ and on the CP-violating ratio $\epe$. Among these 
processes only the decay $B\to X_s\gamma$ has been so far considered in the 
ACD model in the literature \cite{AGDEWU}. As we will see our result 
is consistent with the one obtained by these authors but differs in details.

Of particular interest are the decays $B\to X_s\gamma$ and 
$B\to X_s\mu^+\mu^-$ as the hadronic uncertainties in these decays are 
moderate and the experimental branching ratios are found to be in the 
ballpark of the SM expectations. Thus in principle some constraints on the 
compactification scale $1/R$ could be obtained from these decays. On the 
other hand the upper bound on $K_L\to \pi^0e^+e^-$ is still two orders of 
magnitude above the predictions of the SM and the decay 
$B\to X_s~ {\rm gluon}$ and the ratio $\epe$ are subject 
to very large hadronic uncertainties. Still, as we will see below, 
our results for $B\to X_s~ {\rm gluon}$ and $\epe$ could turn out to be 
important in the future.

The effects of the KK modes on other processes of interest have been 
investigated in a number of papers. In \cite{appelquist:01,APYE} their impact 
on the precision electroweak observables assuming 
a light Higgs ($ m_H \le 250~\gev$) and a heavy Higgs led to the lower bound 
$1/R\ge 300\gev$ and $1/R\ge 250\gev$, respectively. Vacuum stability in a
simplified ACD model has been examined in \cite{Bucci:2003fk}.

Subsequent analyses of the 
anomalous magnetic moment \cite{APDO},
 and of the  
$Z\to b\bar b$ vertex \cite{Santamaria}
have shown the consistency of the ACD 
model with the data for $1/R\ge 300\gev$. The latter calculation has been 
confirmed by us \cite{BSW02}.
 
The scale of $1/R$ as low as $300\gev$ would lead to an exciting
phenomenology in the next generation of colliders 
\cite{COLL0,COLL1,COLL2,COLL3,COLL4}.
Moreover the cosmic relic density 
of the lightest KK 
particle as a dark matter candidate turned out to be of the right order of
magnitude \cite{SETA}. The related experimental signatures have been 
investigated in \cite{DARK}.
In the present paper, as in \cite{BSW02}, 
in order to be more general we will include the results for 
$1/R=200\gev$ that is only slightly below the lowest value of
$1/R=250\gev$ allowed by electroweak precision data.

As our analysis of \cite{BSW02} shows, the ACD model with one extra dimension
has a number of interesting properties from
the point of view of FCNC processes 
discussed here. These are: 

\begin{itemize}
\item
GIM mechanism \cite{GIM} that improves significantly the convergence of 
the sum over the KK modes 
corresponding to the top quark, removing simultaneously to an excellent 
accuracy the contributions of the KK modes corresponding to lighter 
quarks and leptons. This feature removes the sensitivity of the calculated
branching ratios to the scale $M_s\gg 1/R$ at which the higher dimensional 
theory becomes non-perturbative and at which the towers of the KK particles 
must be cut off in an appropriate way. This should be contrasted with 
models with fermions localized on the brane, in which the KK parity is not 
conserved and the sum over the KK modes diverges. 
In these models the results are sensitive to $M_s$ and the KK 
effects in $\Delta M_{s,d}$ are significantly larger \cite{OLPASA} than 
found in \cite{BSW02}. We expect similar behaviour in the processes
considered here.
\item
The low energy effective Hamiltonians are governed by local operators 
already present  in the SM. As flavour violation and CP violation in 
this model is entirely governed by the CKM matrix, the ACD model belongs 
to the class of the so-called models with minimal flavour violation (MFV) 
as defined in \cite{UUT}. This has automatically the following important 
consequence for the FCNC processes considered in \cite{BSW02} and in 
the present paper:
the impact of the KK modes on the processes in question amounts 
only to the modification of the Inami-Lim one-loop functions
\cite{IL}. 
\item
Thus in the case of the processes considered in \cite{BSW02}
these modifications have to be made in the function $S$ \cite{BSS} in 
the case of $\Delta M_{d,s}$
and of the parameter $\varepsilon_K$ and in the functions $X$ and $Y$
\cite{PBE0}  in the case of the rare decays analyzed there. In order to study 
the decays
$B\to X_s\gamma$, $B\to X_s~ {\rm gluon}$, $B\to X_s\mu^+\mu^-$ and 
$K_L\to \pi^0e^+e^-$ and the CP-violating ratio $\epe$ the KK contributions 
to new short distance functions have to be computed. These are 
the functions $D$ (the $\gamma$ penguins), $E$ (gluon penguins), 
$D'$ ($\gamma$-magnetic penguins) and $E'$ (chromomagnetic 
penguins).
In the ACD model these functions 
depend only on $\mt$ and the single new parameter, the compactification 
radius $R$.
\end{itemize}

Our paper is organized as follows. In section 2, we summarize those 
ingredients of the ACD model that are relevant for our analysis. In 
particular, we give in appendix~\ref{Feynmanrules} 
the set of the relevant Feynman rules
involving photons and gluons that have not been given so far in the 
literature. 
Further details can be found in \cite{BSW02,appelquist:01}.
In 
section 3, we calculate the KK contributions to the penguin functions 
$D$, $E$, $D'$ and $E'$.
In section 4, we analyze the decay $B\to X_s\gamma$ and compare our result 
with the one of \cite{AGDEWU}. 
In section 5, we discuss the decay $B\to X_s~{\rm gluon}$.
The corresponding 
analyses of $B\to X_s\mu^+\mu^-$, 
$K_L\to \pi^0e^+e^-$ and $\epe$ are presented in sections  6, 7 and 8, 
respectively.  
In section 9, we summarize our results and give a brief outlook.
As a byproduct of our work we point out a correlation between
the zero $\hat s_0$ in  the $A_{\rm FB}$ asymmetry in $B\to X_s\mu^+\mu^-$ and 
$Br(B\to X_s\gamma)$ that should be valid in most models with minimal 
flavour violation. We also generalize the background field method to 
five dimensions.

\section{The Five Dimensional ACD Model}
\setcounter{equation}{0}

The five dimensional ACD model \cite{appelquist:01} has been described
already
in detail in our previous paper \cite{BSW02} where we have given all
Feynman rules necessary for the calculations considered there. There
are two new features in this work. The first one is the addition of
QCD in 5 dimensions which is straightforward to formulate and has
already been discussed in some detail in \cite{Muck:2001yv}. The other
one is the background field method described in appendix
\ref{backgroundfieldmethod} which is used in the off-shell matching
calculation.
The additional Feynman rules that we need in the present paper, namely
those involving photons and gluons, are collected in appendix
\ref{Feynmanrules}.

Here we only recall the most important features of the ACD model.
The fifth dimension is compactified on the orbifold $S^1/Z_2$ to
reproduce chiral fermions in 4 dimensions. In addition to the ordinary
particles of the SM, denoted as zero modes ($n=0$), there are infinite
towers of KK modes ($n\ge 1$). There is one such tower for each SM
boson and two for each SM fermion, while there also exist physical
scalars $a^0\kk n$ and $a^\pm\kk n$ with ($n\ge 1$) that do not have
any zero mode partners. The masses of the KK particles are universally
given by
\be
   m\kk n^2 = m_0^2+\norsq.
\ee
Here $m_0$ is the mass of the zero mode, as $M_W$, $M_Z$, $m_t$. For
$a^0\kk n$ and $a^\pm\kk n$ this is $M_Z$ and $M_W$, respectively. The
compactification radius $R$ is the only additional parameter
relative to the SM.
To a quark $SU(2)_L$ doublet, there correspond the KK modes $\mcQ$,
whereas the partners of the right-handed singlets are $\mcU$ and
$\mcD$.

As the KK modes contribute to the process considered here first at the
one loop level, the natural variables that enter the Inami-Lim
functions are
\be\label{mindef}
   x\kkx i n =\frac{m\kkx in^2}{M\kkx Wn^2}
\ee
with $m\kkx in$ denoting the masses of the fermionic KK modes and
$M\kkx Wn$ the masses of the $W$ boson KK modes. However, in
phenomenological applications it is more useful to work with the
variables $x_t$ and $x_n$ defined through
\be\label{xtxn}
   x_t = \frac{m_t^2}{M_W^2},\qquad  x_n = \frac{m_n^2}{M_W^2},\qquad 
m_n=\nor
\ee
than with $x\kkx in$.

\section{Penguin Diagrams in the ACD Model}
\setcounter{equation}{0}
\subsection{Preliminaries}
The rare decays in the ACD model considered here are governed as in the 
SM by various penguin diagrams. The SM contributions to $\Delta F=1$ 
box diagrams are subleading but non-negligible. On the other hand the KK
contributions to the latter diagrams are tiny \cite{BSW02} and can be 
neglected.

The penguin vertices including electroweak counter terms can be conveniently
expressed in terms of the functions
$C$ ($Z^0$ penguins), $D$ ($\gamma$ penguins), $E$ (gluon penguins), 
$D'$ ($\gamma$-magnetic penguins) and $E'$ (chromomagnetic 
penguins). In the  
't Hooft--Feynman gauge for the $W^\pm$ and $G^\pm$ propagators they are  
given as follows:
\begin{equation}\label{ZRULE}
 \bar s Z d~ =~i \lambda_t {{G_{\rm F}}\over{\sqrt 2}} {{g_2}\over{2\pi^2}} 
   {{M^2_W}\over{\cos\theta_{w}}} C(x_t,1/R) \bar s \gamma_\mu 
   (1-\gamma_5)d
\end{equation}
\begin{equation}\label{Ddef}
 \bar s\gamma d~ =~- i\lambda_t {{G_{\rm F}}\over{\sqrt 2}} {{e}\over{8\pi^2}}
   D(x_t,1/R) \bar s (q^2\gamma_\mu - q_\mu \not\!q)(1-\gamma_5)d 
\end{equation}
\begin{equation}\label{Edef}
 \bar s G^a d~ =~ -i\lambda_t{{G_{\rm F}}\over{\sqrt 
2}} {{g_s}\over{8\pi^2}}
   E(x_t,1/R) \bar s_{\alpha}(q^2\gamma_\mu - q_\mu \not\!q)
(1-\gamma_5)T^a_{\alpha\beta}d_\beta 
\end{equation}
\begin{equation}\label{MGP}
 \bar s \gamma' b~ =~i\bar\lambda_t {{G_{\rm F}}\over{\sqrt 2}} {{e}\over
   {8\pi^2}} D'(x_t,1/R) \bar s \lbrack i\sigma_{\mu\lambda} q^\lambda
   \lbrack m_b (1+\gamma_5) \rbrack\rbrack b
\end{equation}
\begin{equation}\label{FRF}
 \bar s G'^a b~ =~ 
i\bar\lambda_t{{G_{\rm F}}\over{\sqrt 2}}{{g_s}\over{8\pi^2}}
   E'(x_t,1/R)\bar s_{\alpha} \lbrack i\sigma_{\mu\lambda} q^\lambda
   \lbrack m_b (1+\gamma_5) \rbrack\rbrack T^a_{\alpha\beta} b_\beta \,,
\end{equation}
where $G_F$ is the Fermi constant, $\theta_w$ is the weak mixing angle and 
\be
\lambda_t=V^*_{ts}V_{td},\quad\quad \bar\lambda_t=V^*_{ts}V_{tb}~. 
\ee
In these vertices
$q_\mu$ is the {\it outgoing} gluon or photon momentum
and $T^a$ are colour matrices.
The last two vertices involve on-shell photon
and gluon, respectively.
We have set $m_s=0$ in these vertices. 

Each function in (\ref{ZRULE})--(\ref{FRF}) is given by 
\be\label{FACD}
F(x_t,1/R)=F_0(x_t)+\sum_{n=1}^\infty F_n(x_t,x_n), \qquad F=C,D,E,D',E',
\ee
with $x_n$ defined in (\ref{xtxn}). 
The functions $F_n(x_t,x_n)$ in (\ref{FACD}) are defined through
\be\label{CN}
F_n(x_t,x_n)= G(x_{t(n)})-G(x_{u(n)}),
\ee
with $x\kkx in$ defined in (\ref{mindef}) and the functions
$G(x_{t(n)})$ and $G(x_{u(n)})$ representing the contributions of the 
$\mcQ\kkx tn$, $\mcU\kkx tn$ and $\mcQ\kkx un$, $\mcU\kkx un$ modes,
respectively.

The functions
$F_0(x_t)$ result from the penguin diagrams in the SM and 
the sum represents the KK contributions to the relevant penguin diagrams.
$F_0(x_t)$ were calculated by various authors, in
particular by Inami and Lim \cite{IL}.  
They are given explicitly as follows:
\begin{equation}\label{C0xt}
C_0(x_t)={x_t\over 8}\left[{{x_t-6}\over{x_t-1}}+{{3x_t+2}
\over{(x_t-1)^2}}\;\ln x_t\right]~, 
\end{equation}
\begin{equation}
D_0(x_t)=-{4\over9}\ln x_t+{{-19x_t^3+25x_t^2}\over{36(x_t-1)^3}}
+{{x_t^2(5x_t^2-2x_t-6)} \over{18(x_t-1)^4}}\ln x_t~,
\end{equation}
\begin{equation}\label{E0}
E_0(x_t)=-{2\over 3}\ln x_t+{{x_t^2(15-16x_t+4x_t^2)}\over{6(1-x_t)^4}}
\ln x_t+{{x_t(18-11x_t-x_t^2)} \over{12(1-x_t)^3}}~,
\end{equation}
\begin{equation}
D'_0(x_t)= -{{(8x_t^3 + 5x_t^2 - 7x_t)}\over{12(1-x_t)^3}}+ 
          {{x_t^2(2-3x_t)}\over{2(1-x_t)^4}}\ln x_t~,
\end{equation}
\begin{equation}
E'_0(x_t)=-{{x_t(x_t^2-5x_t-2)}\over{4(1-x_t)^3}} + {3\over2}
{{x_t^2}\over{(1 - x_t)^4}} \ln x_t~.
\end{equation}

The $Z^0$ penguin functions $C_n(x_t,x_n)$ have been calculated in 
\cite{BSW02} with the result
\be\label{CFIN}
C_n(x_t,x_n)=\frac{x_t}{8 (x_t-1)^2}
\left[x_t^2-8 x_t+7+(3 +3 x_t+7 x_n-x_t
x_n)\ln \frac{x_t+x_n}{1+x_n}\right]. 
\ee
In the present paper we will calculate the remaining functions 
$D_n(x_t,x_n)$, $E_n(x_t,x_n)$, 
$D'_n(x_t,x_n)$ and $E'_n(x_t,x_n)$.

\subsection{General Structure of the Calculation}
\begin{figure}[hbt]
  \centering
  \subfigure[]{
    \includegraphics[scale=0.9]{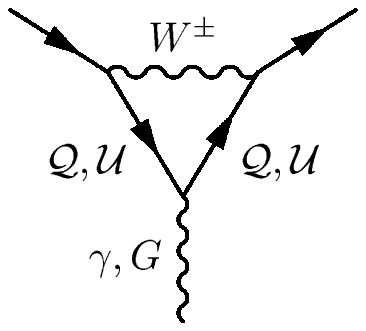}
    }
  \hspace{1.5cm}
  \subfigure[]{
    \includegraphics[scale=0.9]{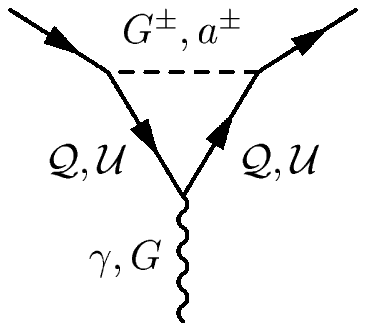}
    }
  \hspace{1.5cm}
  \subfigure[]{
    \includegraphics[scale=0.9]{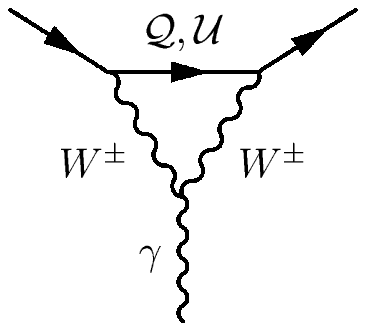}
    }
  \hspace{1.5cm}
  \subfigure[]{
    \includegraphics[scale=0.9]{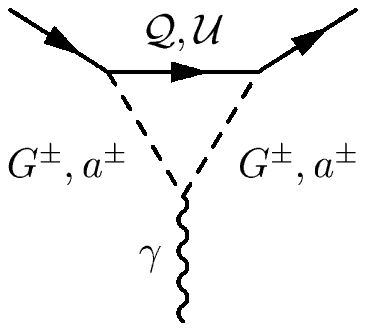}
    }
  \hspace{1.5cm}
  \subfigure[]{
    \includegraphics[scale=0.9]{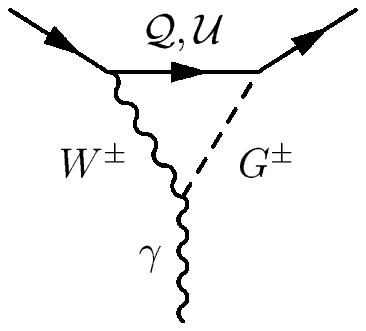}
    }
  \hspace{1.5cm}
  \subfigure[]{
    \includegraphics[scale=0.9]{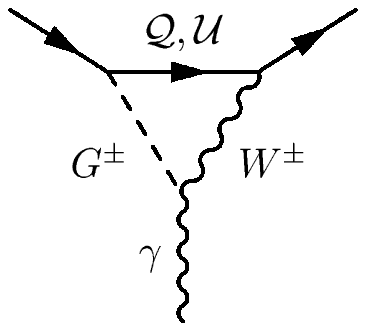}
    }
  \caption[]{\small\label{penguindiagrams} Penguin diagrams contributing to
  $F_n(x_t,x_n)$.}
\end{figure}

The function $F_n(x_t,x_n)$ with $F=D,E,D',E'$ can be found by
calculating  
the vertex diagrams in fig.~\ref{penguindiagrams}.
Contrary to the $Z^0$ penguins where one has to add an electroweak 
counter term as discussed in \cite{BB1}, this is not necessary for the
$\gamma$ and gluon penguins. Here the counter terms are only
formally used to render zero the coefficients of the dimension 4
operators $\bar s\fslash A P_Lq$ and $\bar s\mathbf T^a\fslashl G^a
P_Lq$ with $q=d,b$. This is a consequence of gauge 
invariance when the quark fields are set on-shell
with the renormalization condition given in \cite{Denner:kt}.

In contrast to the calculation of the $Z^0$-vertex the
external momenta in fig.~\ref{penguindiagrams} and 
the masses of external quarks cannot be neglected.

The calculation of the functions $F$ was done in two different
ways. The common starting point is the effective Hamiltonian at the
matching scale $\mu_W$
\be\label{HeffmuW}
  {\cal H}_{\rm eff}(b\to s\gamma,G) = - \frac{G_{\rm F}}{\sqrt{2}}
  V_{ts}^* V_{tb} \left[ \sum_{i=30}^{36} C_i(\mu_W) Q_i +
    C_{7\gamma}(\mu_W) Q_{7\gamma} +C_{8G}(\mu_W) Q_{8G} \right]\,,
\ee
with the operators \cite{Bobeth}
\begin{align}
  \label{Q7gamma}
  Q_{7\gamma} &= \frac{e}{4\pi^2}m_b(\bar
  s\sigma_{\mu\nu}P_Rb)F^{\mu\nu},\\ 
  Q_{8G} &= \frac{g_s}{4\pi^2}m_b(\bar
  s\sigma_{\mu\nu}P_R \mathbf T^a b)G^{a,\mu\nu},\\
  Q_{30} &= \frac{i}{4\pi^2}M_W^2(\bar s\fslashl D P_Lb),\\
  Q_{31} &= \frac{g_s}{4\pi^2}(\bar s P_R\gamma_\mu \mathbf T^ab)D_\nu
    G^{a,\mu\nu}+Q_4,\\ 
  Q_{32} &= \frac{1}{4\pi^2}m_b(\bar sP_R \fslashl D\fslashl Db),\\
  Q_{33} &= \frac{i}{4\pi^2}(\bar sP_R\fslashl D\fslashl D\fslashl
  Db),\\ 
  Q_{34} &= \frac{ig_s}{4\pi^2}\bar sP_R\left[\fslashl{\overleftarrow
      D}\sigma_{\mu\nu}\mathbf T^a - \mathbf
    T^a\sigma_{\mu\nu}(\fslashl D+im_b)\right] bG^{a,\mu\nu},\\ 
  Q_{35} &= \frac{ie}{4\pi^2}\bar sP_R\left[\fslashl{\overleftarrow
      D}\sigma_{\mu\nu} - \sigma_{\mu\nu}(\fslashl D+im_b)\right]
  bF^{\mu\nu},\\  
  \label{Q36}
  Q_{36} &= \frac{e}{4\pi^2}(\bar s P_R\gamma_\mu b)\partial_\nu
  F^{\mu\nu}-Q_9. 
\end{align}

The covariant derivative in the effective Lagrangian is defined
as\footnote{On the effective side we use a different sign convention
  for the couplings $e$ and $g_s$ compared to the Feynman rules in
  this paper and in \cite{BSW02}.}
\be
  D_\mu\psi = (\partial_\mu +ieQ_\psi A_\mu +ig_s\mathbf
  T^aG^a_\mu)\psi.
\ee
In the operators $Q_{34}$ and $Q_{35}$ it acts only on the spinors,
but not on the field strength tensors $F^{\mu\nu}$ and
$G^{a,\mu\nu}$. 

In this off-shell operator basis we have omitted the four-quark
operators as they are not relevant for the calculation of the
functions F. They can be found in \cite{Bobeth}, in particular the
operators $Q_4$ and $Q_9$ that appear in the definition of $Q_{31}$
and $Q_{36}$.
The effective vertices (\ref{Ddef})-(\ref{FRF}) correspond to
the operators $Q_{36}$, $Q_{31}$, $Q_{7\gamma}$ and $Q_{8G}$,
respectively. 

In the first method the equations of motion of the quark fields and
the relation $q^2=0$ are
applied both on the full and on the effective side of the matching
amplitude. After renormalization the amplitude on the effective side
is 
\be
  \mcA^{b\rightarrow
    s\gamma}_\text{eff}=\bar\lambda_t\frac{G_F}{\sqrt{2}}\frac{ie}{8\pi^2}  
  \bar s\left[D'(2 T_1+ T_2)-D T_2\right]b 
\ee
with $T_1=m_b\fslash q\gamma_\mu P_R$ and $T_2=-2m_bq_\mu P_R$.

The full side was evaluated with the Feynman diagrams (1)-(6) in
fig.~\ref{penguindiagrams} in 't~Hooft-Feynman gauge. The vertices
$A\kk 0 W^\pm\kk n a^\mp\kk n$ and $A\kk 0 G^\pm\kk n a^\mp\kk n$
are zero, hence there are no diagrams with these vertices to be
considered. 
The result can be written as
\be
  \mcA^{b\rightarrow
    s\gamma}_\text{full}=\bar\lambda_t\frac{G_F}{\sqrt{2}}\frac{ie}{8\pi^2} 
  \bar s\left[H_1 T_1 + H_2 T_2\right]b,
\ee
yielding
\be
  D=\frac{1}{2}H_1 - H_2,\qquad D'=\frac{1}{2}H_1.
\ee

For the functions $E$ and $E'$ the procedure is analogous. As the
gluon couples only to the quarks the relevant diagrams in this case
are (1) and (2). 

The second method uses the off-shell amplitude for the matching,
i.e. no equations of motion are applied. In order to
maintain gauge invariance the method of background fields
\cite{Abbott} is used. Here the matching is done with the full set of
operators (\ref{Q7gamma})-(\ref{Q36}).

For the full side the diagrams (1)-(4) in fig.~\ref{penguindiagrams}
were evaluated in 't~Hooft-Feynman gauge with the Feynman rules for
the background fields $\hat\gamma$ and $\hat G$ given in
appendix~\ref{Feynmanrules}.
As a virtue of the background field (BF) method, the diagrams (5) and
(6) do not show up because the vertex $\hat A\kk 0 W^\pm\kk
n G^\mp\kk n$ vanishes.

The effective side can be directly matched to the full side. Comparing
(\ref{Ddef})-(\ref{FRF}) with (\ref{HeffmuW}) we can read off
\begin{align}
  D = -C^{(0)}_{36}(\mu_W),&\qquad D' = -2C^{(0)}_{7\gamma}(\mu_W),\\
  E = -C^{(0)}_{31}(\mu_W),&\qquad E' = -2C^{(0)}_{8G}(\mu_W),
\end{align}
where ``(0)" indicates the leading coefficients without QCD corrections.
We found the same results for the functions $F$ with both methods.

\subsection{The functions \boldmath{$D$}, \boldmath{$E$} \boldmath{$D'$} 
\boldmath{$E'$} and \boldmath{$Z$}} 
Adding up the contributions from the diagrams of fig.~\ref{penguindiagrams}
we find
\bea\label{DFIN}
D_n(x_t,x_n)&=& \frac{x_t \left( 35 + 8 x_t - 19 x_t^2 + 
       6 x_n^2 \left( 10 - 9 x_t + 3 x_t^2 \right)  + 
       3 x_n \left( 53 - 58 x_t + 21 x_t^2 \right)  \right) 
     }{108 {\left(x_t - 1 \right) }^3}\nonumber\\ &&+ 
  \frac{1}{6}\left( 4 - 2 x_n + 4 x_n^2 + x_n^3 \right)  
     \ln \frac{x_n}{1 + x_n}
 \nonumber\\ && \hspace{-1.5cm} - 
  \frac{1}{18 
     {\left(x_t - 1 \right) }^4}\left( 12 - 38 x_t + 54 x_t^2 - 27 x_t^3 + 
       3 x_t^4 + x_n^3 \left( 3 + x_t \right)  + 
       3 x_n^2 \left( 4 - x_t + x_t^2 \right) \right. \nnl &&\hspace{-1.5cm} + \left.
       x_n \left( -6 + 42 x_t - 33 x_t^2 + 
          9 x_t^3 \right)  \right)  
     \ln \frac{x_n + x_t}{1 + x_n},
\eea
\bea\label{EFIN}
E_n(x_t,x_n)&=&- \frac{ x_t\left( 35 + 8 x_t - 19 x_t^2 + 
         6 x_n^2 \left( 10 - 9 x_t + 3 x_t^2 \right)  + 
         3 x_n \left( 53 - 58 x_t + 21 x_t^2 \right) 
         \right)   }{36 {\left(x_t - 1 \right) }^3}  \nnl&&-
  \frac{1}{2}\left( 1 + x_n \right)  
     \left( -2 + 3 x_n + x_n^2 \right)  
     \ln \frac{x_n}{1 + x_n}\nonumber\\ &&\hspace{-1.5cm}   + 
  \frac{\left( 1 + x_n \right)  
     \left( -6 + 19 x_t - 9 x_t^2 + 
       x_n^2 \left( 3 + x_t \right)  + 
       x_n \left( 9 - 4 x_t + 3 x_t^2 \right)  \right)  
     }{6 
     {\left(x_t - 1 \right) }^4}\ln \frac{x_n + x_t}{1 + x_n},
\eea
\bea\label{DPFIN}
D'_n(x_t,x_n)&=&\frac{x_t \left( -37 + 44 x_t + 17 x_t^2 + 
       6 x_n^2 \left( 10 - 9 x_t + 3 x_t^2 \right)  - 
       3 x_n \left( 21 - 54 x_t + 17 x_t^2 \right)  \right) }{36 
     \left( -1 + x_t \right)^3} \nnl
&&+ \frac{x_n 
     \left( 2 - 7 x_n + 3 x_n^2 \right) }{6} \ln \frac{x_n}{1 + x_n}\nnl
&&\hspace{-2cm} - 
  \frac{\left( -2 + x_n + 3 x_t \right)  
     \left( x_t + 3 x_t^2 + x_n^2 \left( 3 + x_t \right)  - 
       x_n \left( 1 + \left( -10 + x_t \right)  x_t \right)  \right)  
    }{6 \left( -1 + x_t \right)^4} \ln \frac{x_n + x_t}{1 + x_t},
\eea
\bea\label{EPFIN}
E'_n(x_t,x_n)& =&\frac{x_t \left( -17 - 8 x_t + x_t^2 - 
       3 x_n \left( 21 - 6 x_t + x_t^2 \right)  - 
       6 x_n^2 \left( 10 - 9 x_t + 3 x_t^2 \right) 
       \right) }{12 {\left(x_t - 1 \right) }^3}\nnl && - 
  \frac{1}{2}x_n \left( 1 + x_n \right)  
     \left( -1 + 3 x_n \right)  \ln \frac{x_n}{1 + x_n}\nonumber\\
  &&\hspace{-1.5cm} +
   \frac{\left( 1 + x_n \right)  
     \left( x_t + 3 x_t^2 + 
       x_n^2 \left( 3 + x_t \right)  - 
       x_n \left( 1 + \left( -10 + x_t \right)  x_t \right) 
       \right)  }{2 
     {\left(x_t - 1 \right) }^4}\ln \frac{x_n + x_t}{1 + x_n}.
\eea

\begin{figure}[hbt]
\renewcommand{\thesubfigure}{\space(\alph{subfigure})} 
  \centering 
  \subfigure[]{\psfragscanon
    \psfrag{dacdnum}{$D(x_t,1/R)$}
    \psfrag{rinvrinv}[][]{\shortstack{\\  $R^{-1}$ [GeV] } }
    \resizebox{.35\paperwidth}{!}{\includegraphics[]{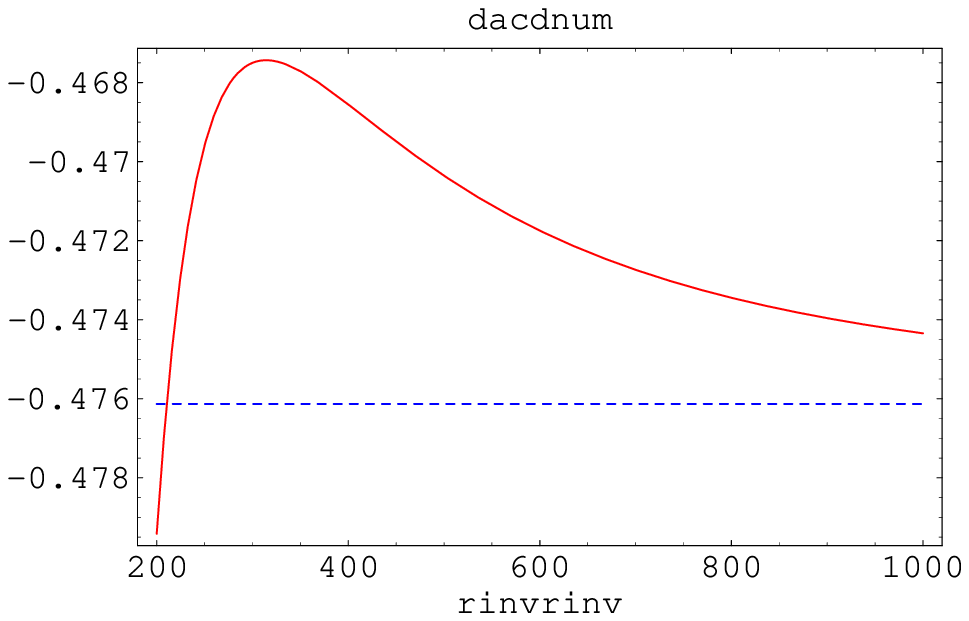}}
    }
  \subfigure[]{\psfragscanon
    \psfrag{dprime}{$D'(x_t,1/R)$}
    \psfrag{rinvrinv}[][]{\shortstack{\\  $R^{-1}$ [GeV] } }
    \resizebox{.35\paperwidth}{!}{\includegraphics[]{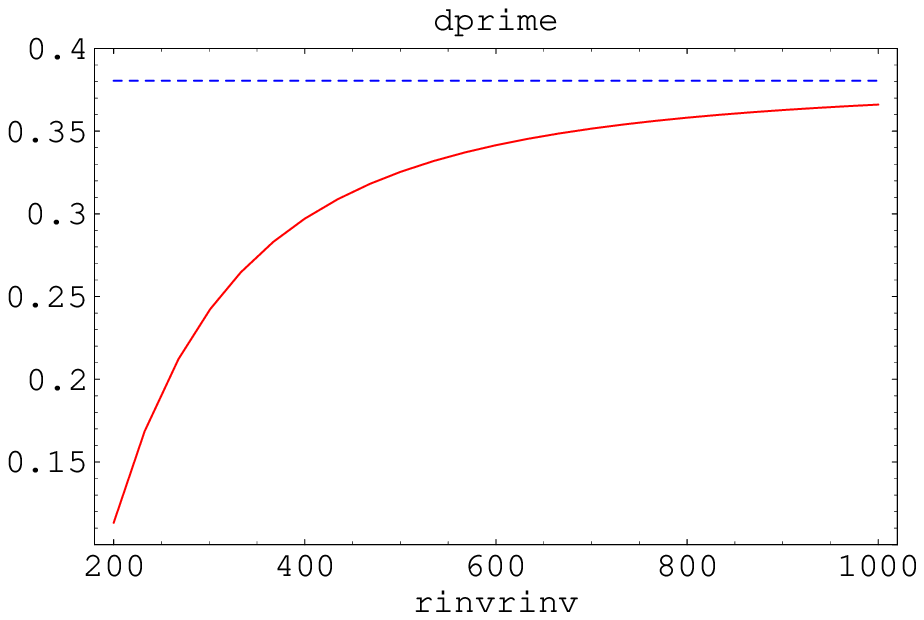}}
    }
  \subfigure[]{\psfragscanon
    \psfrag{eacdnum}{$E(x_t,1/R)$}
    \psfrag{rinvrinv}[][]{\shortstack{\\ $R^{-1}$ [GeV]}  }
    \resizebox{.35\paperwidth}{!}{\includegraphics[]{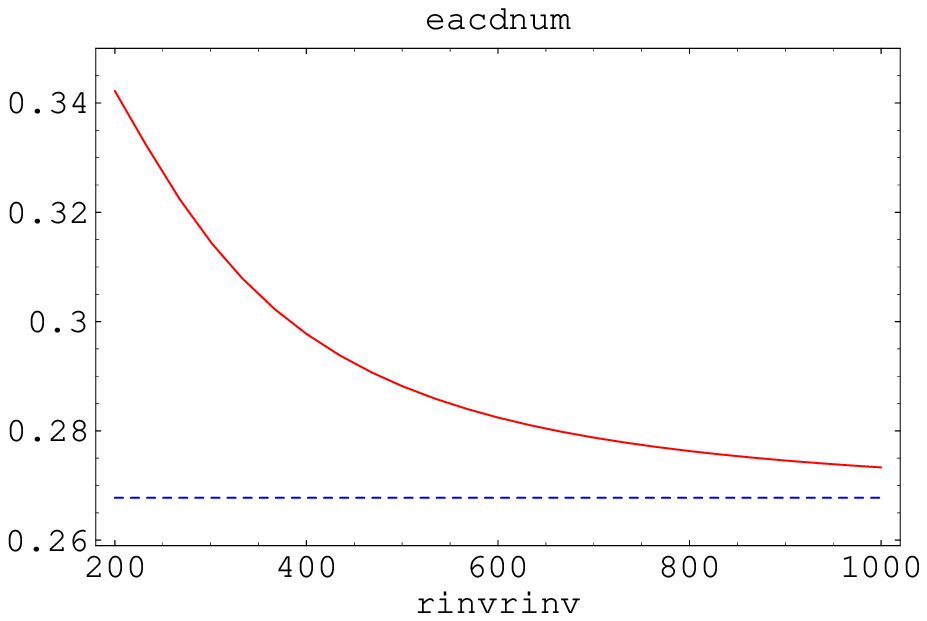}}
    }
  \subfigure[]{\psfragscanon
    \psfrag{eprime}{$E'(x_t,1/R)$}
    \psfrag{rinvrinv}[][]{\shortstack{\\  $R^{-1}$ [GeV] } }
    \resizebox{.35\paperwidth}{!}{\includegraphics[]{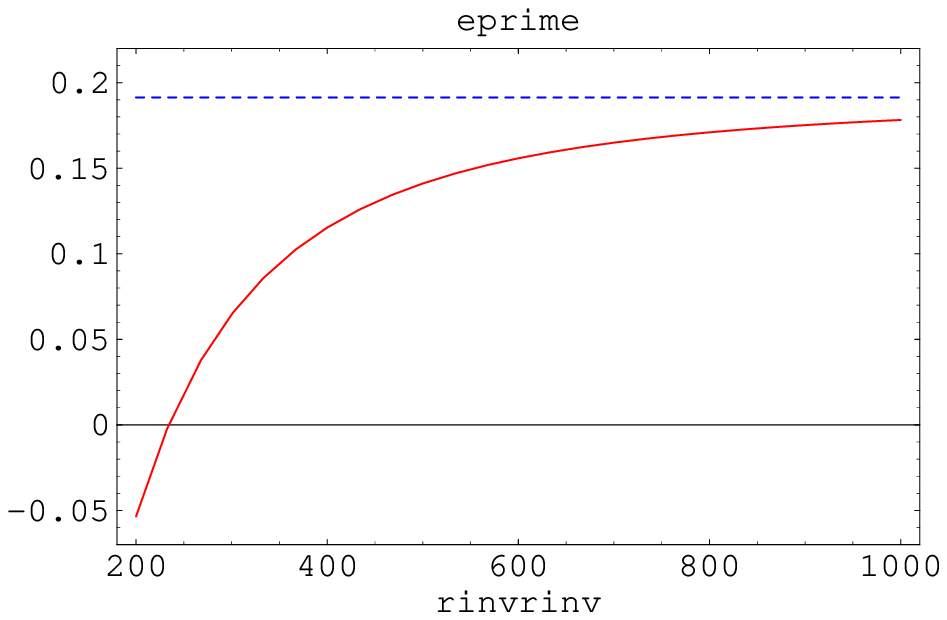}}
    }
  \caption[]{\small\label{EDplot} The functions $F(x_t,1/R)$
    and $F_0(x_t)$ with  $F=D,E,D',E'$ plotted versus $1/R$. The
    constant dashed lines are the SM values.}
\end{figure}

In fig.~\ref{EDplot} we plot the functions $F(x_t,1/R)$ versus 
$1/R$. The impact of the KK modes on the function $D$ is negligible. 
The function $E$ is moderately enhanced but this enhancement plays only 
a marginal role in our phenomenological applications. The most interesting 
are very strong suppressions of $D'$ and $E'$, that for $1/R=300\gev$ amount 
to $36\%$ and $66\%$ relative to the SM values, respectively.


While the $\Delta F=2$ box function $S$ and the penguin functions 
$E$, $D'$ and $E'$ are gauge independent, this is not the case for 
$C$, $D$ and the $\Delta F=1$ box diagram functions 
$B^{\nu\bar\nu}$ and $B^{\mu\bar\mu}$
considered in 
\cite{BSW02}. In phenomenological applications it is more 
convenient to work with gauge independent functions \cite{PBE0}
\begin{equation}\label{xx9} 
X(x_t,1/R)=C(x_t,1/R)+B^{\nu\bar\nu}(x_t,1/R)=X_0(x_t)+\Delta X~, 
\end{equation}
\begin{equation}\label{yyx}
Y(x_t,1/R)  =C(x_t,1/R)+B^{\mu\bar\mu}(x_t,1/R)=Y_0(x_t)+\Delta Y~,
\end{equation}
\begin{equation}\label{zzx}
Z(x_t,1/R)  =C(x_t,1/R)+\frac{1}{4}D(x_t,1/R)=Z_0(x_t)+\Delta Z
\end{equation}
with ($\mt=167\gev$)
\begin{equation}\label{X0}
X_0(x_t)={x_t\over{8}}\;\left[{{x_t+2}\over{x_t-1}} 
+ {{3 x_t-6}\over{(x_t -1)^2}}\; \ln x_t\right]= 1.526 ~,
\end{equation}
\begin{equation}\label{Y0}
Y_0(x_t)={x_t\over8}\; \left[{{x_t -4}\over{x_t-1}} 
+ {{3 x_t}\over{(x_t -1)^2}} \ln x_t\right]= 0.980
\end{equation}
\bea\label{Z0}
Z_0(x_t)&=&-{1\over9}\ln x_t + 
{{18x_t^4-163x_t^3 + 259x_t^2-108x_t}\over {144 (x_t-1)^3}}+
\nonumber\\ 
& &+{{32x_t^4-38x_t^3-15x_t^2+18x_t}\over{72(x_t-1)^4}}\ln x_t=0.679
\eea
summarizing the SM contributions and $\Delta X$, $\Delta Y$ and $\Delta Z$
representing the corrections due to KK modes.
Our analysis of \cite{BSW02} and the negligible KK contributions to 
the function $D$ found here imply to an excellent approximation
\be
\Delta X=\Delta Y=\Delta Z=\sum_{n=1}^\infty C_n(x_t,x_n).
\ee

In table~\ref{XYZ} we give the values of the gauge independent functions 
$S$, $X$, $Y$ calculated in \cite{BSW02} and of $Z$, $E$, $D'$, $E'$ 
calculated 
here for different $1/R$ and $m_t=167\GeV$. 
Our results for the function $S$ have been confirmed in \cite{CHHUKU}.
For completeness we give also 
the values of $C$ and $D$ in 't Hooft--Feynman gauge.
We observe that for $1/R=300\GeV$, the functions $X$, $Y$, $Z$ are enhanced 
by $10\%$, $15\%$ and $23\%$ relative to the SM values, respectively. 
While these enhancements are smaller than the corresponding suppressions of
$D'$ and $E'$, the effect of the latter suppressions  will be softened in 
the relevant branching ratios through sizable additive QCD corrections.

\begin{table}[hbt]
\begin{center}
\begin{tabular}{|c||c|c|c|c|c|c|c||c|c|}\hline
 $1/R~[{\rm GeV}]$  & {$S$} & {$X$}& {$Y$} & {$Z$} & {$E$} & {$D'$} & {$E'$} 
& {$C$} & $D$
 \\ \hline
$ 200$ & $ 2.813 $ &  $1.826 $ &  $1.281$ & $0.990$ &$ 0.342$  & $0.113$  & $-0.053$ & $1.099$  & $-0.479$  
\\ \hline
$ 250$ & $ 2.664 $ &  $1.731 $ &  $1.185$ & $0.893$ & $0.327$  & $0.191$ &$ 0.019$ & $1.003$  & $-0.470$
\\ \hline
$300$  & $ 2.582 $ &  $1.674 $ &  $1.128$ & $0.835$ & $0.315$  & $0.242$  &$ 0.065$ & $0.946$  & $-0.468$
\\ \hline
$400$  & $ 2.500 $ &  $1.613 $ &  $1.067$ & $0.771$ & $0.298$  & $0.297$  &$ 0.115 $ & $0.885$  & $-0.469$
\\ \hline
SM     & $2.398$   &  $ 1.526 $ & $0.980$ & $0.679$ & $0.268$  &$ 0.380$ & $ 0.191$ & $0.798$  & $-0.476$
\\ \hline
\end{tabular}
\end{center}
\caption[]{\small Values for the functions $S$, $X$, $Y$, 
$Z$, $E$, $D'$, $E'$, $C$ and $D$.
\label{XYZ}}
\end{table}

\section{\boldmath{$B\to X_s\gamma$}} 
\setcounter{equation}{0}
\subsection{Preliminaries}
The inclusive $B\to X_s\gamma$ decay has been the subject of very intensive 
theoretical and experimental studies during the last 15 years. On the 
experimental side the world average resulting from the data by CLEO, ALEPH, 
BaBar and Belle reads \cite{Battaglia:2003in}
\be\label{bsgexp}
Br(B\to X_s\gamma)_{E_\gamma> 1.6\GeV}= (3.28^{+0.41}_{-0.36})\cdot 10^{-4}~.
\ee
It agrees well with the SM result \cite{Gambino:2001ew,Buras:2002tp}
\be\label{bsgth}
Br(B\to X_s\gamma)^{\text{SM}}_{E_\gamma> 1.6\GeV}
= (3.57\pm 0.30)\cdot 10^{-4}~.
\ee
The most recent reviews summarizing the theoretical status can be found 
in \cite{AJBMIS,ALIMIS,HURTH}. The purpose of this section is 
the calculation of
the impact of the KK modes on the branching ratio in question. 

As the NLO QCD corrections to $Br(B\to X_s\gamma)$ within the SM 
are very important, it is mandatory to include them also in the ACD 
model. Fortunately, the QCD corrections for scales $\mu<1/R$ are in the 
ACD model precisely the same as in the SM and the only elements of the 
NLO QCD corrections to  $Br(B\to X_s\gamma)$ within the ACD model that 
are missing, are the $\ord(\alpha_s)$ corrections to the functions 
$D'$ and $E'$. The experience with such corrections within the SM and 
MSSM at low $\tan\beta$ makes us to expect that they are significantly 
smaller than the renormalization group effects for $\mu< \mw$ and the 
$\ord(\alpha_s)$ corrections at scales $\ord(m_b)$.
Therefore we are confident that it is a good approximation to neglect these 
corrections in the case of the KK contributions. 

Now, the NLO formulae for the relevant Wilson coefficients and the 
$Br(B\to X_s\gamma)$ are very complicated and not transparent. Therefore we 
will first discuss the impact of the KK contributions at the LO level. 
Subsequently we will present the numerical results of an NLO analysis that
uses the formulae of \cite{Gambino:2001ew} modified by the KK contributions to 
the functions $D'$ and $E'$ calculated in the previous section.

\subsection{Effective Hamiltonian}
The effective Hamiltonian for $B\to X_s\gamma$ at scales 
$\mu_b={\cal O}(m_b)$
is given by
\begin{equation} \label{Heff_at_mu}
{\cal H}_{\rm eff}(b\to s\gamma) = - \frac{G_{\rm F}}{\sqrt{2}} V_{ts}^* V_{tb}
\left[ \sum_{i=1}^6 C_i(\mu_b) Q_i + C_{7\gamma}(\mu_b) Q_{7\gamma}
+C_{8G}(\mu_b) Q_{8G} \right]\,,
\end{equation}
where in view of $\mid V_{us}^*V_{ub} / V_{ts}^* V_{tb}\mid < 0.02$
we have neglected the term proportional to $V_{us}^* V_{ub}$.
Here $Q_1....Q_6$ are the usual four-quark operators whose
explicit form can be found in \cite{AJBLH}. 
The remaining two operators,
characteristic for this decay, are the {\it magnetic--penguins}
\begin{equation}\label{O6B}
Q_{7\gamma}  =  \frac{e}{8\pi^2} m_b \bar{s}_\alpha \sigma^{\mu\nu}
          (1+\gamma_5) b_\alpha F_{\mu\nu},\qquad            
Q_{8G}     =  \frac{g_s}{8\pi^2} m_b \bar{s}_\alpha \sigma^{\mu\nu}
   (1+\gamma_5)T^a_{\alpha\beta} b_\beta G^a_{\mu\nu}  
\end{equation}
originating in the diagrams of fig.~\ref{penguindiagrams} with on-shell
photon and gluon, respectively.

The coefficients $C_i(\mu_b)$ in (\ref{Heff_at_mu}) can be calculated
by using
\begin{equation}
 C(\mu_b)= \hat U_5(\mu_b,\mu_W) C(\mu_W)
\end{equation}
Here $ \hat U_5(\mu_b,\mu_W)$ with $\mu_W=\ord(M_W)$ is the $8\times 8$ 
evolution matrix that is known including NLO corrections \cite{CHMIMU}.
As stated before, we will first estimate the effect of the 
KK modes on $C_{7\gamma}(\mu_b)$ and $C_{8G}(\mu_b)$ in the leading
logarithmic (LO) approximation. In this case generalizing the SM formulae
in \cite{BMMP:94} to the ACD model we obtain
\begin{eqnarray}
\label{C7eff}
C_{7\gamma}^{(0)eff}(\mu_b) & = & 
\eta^\frac{16}{23} C_{7\gamma}^{(0)}(\mu_W) + \frac{8}{3}
\left(\eta^\frac{14}{23} - \eta^\frac{16}{23}\right) C_{8G}^{(0)}(\mu_W) +
 C_2^{(0)}(\mu_W)\sum_{i=1}^8 h_i \eta^{a_i},
\\
\label{C8eff}
C_{8G}^{(0)eff}(\mu_b) & = & 
\eta^\frac{14}{23} C_{8G}^{(0)}(\mu_W) 
   + C_2^{(0)}(\mu_W) \sum_{i=1}^8 \bar h_i \eta^{a_i},
\end{eqnarray}
with
\begin{eqnarray}
\eta & = & \frac{\as(\mu_W)}{\as(\mu_b)}, 
\end{eqnarray}
and 
\begin{equation}\label{c2}
C^{(0)}_2(\mu_W) = 1, \qquad                               
C^{(0)}_{7\gamma} (\mu_W) = -\frac{1}{2} D'(x_t,1/R), \qquad
C^{(0)}_{8G}(\mu_W) = -\frac{1}{2} E'(x_t,1/R)
\end{equation}
and all remaining coefficients equal zero at $\mu=\mu_W$. 
The superscript ``0" indicates the LO approximation. 
The ``effective" coefficients in (\ref{C7eff}) and
(\ref{C8eff}) are introduced in place of 
$C_{7\gamma}(\mu_b)$ and $C_{8G}(\mu_b)$ in order to keep the LO
Wilson coefficients renormalization scheme independent as discussed in
\cite{BMMP:94}. 
The functions 
$D'(x_t,1/R)$ and $E'(x_t,1/R)$ have been calculated in section 3.
Finally the values of $a_i$, $h_i$ and $\bar h_i$ are given 
in table \ref{tab:akh}.

\begin{table}[htb]

\begin{center}
\begin{tabular}{|r|r|r|r|r|r|r|r|r|}
\hline
$i$ & 1 & 2 & 3 & 4 & 5 & 6 & 7 & 8 \\
\hline
$a_i $&$ \frac{14}{23} $&$ \frac{16}{23} $&$ \frac{6}{23} $&$
-\frac{12}{23} $&$
0.4086 $&$ -0.4230 $&$ -0.8994 $&$ 0.1456 $\\
$h_i $&$ 2.2996 $&$ - 1.0880 $&$ - \frac{3}{7} $&$ -
\frac{1}{14} $&$ -0.6494 $&$ -0.0380 $&$ -0.0185 $&$ -0.0057 $\\
$\bar h_i $&$ 0.8623 $&$ 0 $&$ 0 $&$ 0
 $&$ -0.9135 $&$ 0.0873 $&$ -0.0571 $&$ 0.0209 $\\
\hline
\end{tabular}
\end{center}
\caption[]{Magic Numbers.
\label{tab:akh}}
\end{table}

Using the leading $\mu_b$-dependence of $\as$ in an effective five quark 
theory
\begin{equation} 
\as^{(5)}(\mu_b) = \frac{\as(\mz)}{1 
- \beta_0 \frac{\as(\mz)}{2\pi} \, \ln(\mz/\mu_b)}, \qquad 
\beta_0=\frac{23}{3}~,
\label{eq:asmumz}
\end{equation} 
and setting $\as^{(5)}(\mz) = 0.118$ we find the results in 
table \ref{tab:c78effnum}.

\begin{table}[htb]
\begin{center}
\begin{tabular}{|c||c|c||c|c||c|c|}
\hline
& \multicolumn{2}{c||}{$\mu_b = 2.5\gev$} &
  \multicolumn{2}{c||}{$\mu_b = 5.0\gev$} &
  \multicolumn{2}{c| }{$\mu_b = 7.5\gev$} \\
\hline
$1/R [\gev]$ & 
$C^{(0){\rm eff}}_{7\gamma}$ & $C^{(0){\rm eff}}_{8G}$ &
$C^{(0){\rm eff}}_{7\gamma}$ & $C^{(0){\rm eff}}_{8G}$ &
$C^{(0){\rm eff}}_{7\gamma}$ & $C^{(0){\rm eff}}_{8G}$ \\
\hline
200 &$ -0.236 $&$ -0.076 $&$ -0.192 $&$ -0.053 $&$ -0.169 $&$ -0.040 $\\
250 &$ -0.264 $&$ -0.100 $&$ -0.223 $&$ -0.079 $&$ -0.201 $&$ -0.068 $\\
300 &$ -0.282 $&$ -0.114 $&$ -0.242 $&$ -0.096 $&$ -0.221 $&$ -0.086 $\\
400 &$ -0.301 $&$ -0.131 $&$ -0.264 $&$ -0.114 $&$ -0.244 $&$ -0.105 $\\
SM  &$ -0.331 $&$ -0.156 $&$ -0.296 $&$ -0.142 $&$ -0.278 $&$ -0.135 $\\
\hline
\end{tabular}
\end{center}
\caption[]{Wilson coefficients $C^{(0){\rm eff}}_{7\gamma}$ and 
$C^{(0){\rm eff}}_{8G}$ in LO
for $\mt = 167 \gev$ as functions of $1/R$ and various values of $\mu_b$.
\label{tab:c78effnum}}
\end{table}

We observe a sizable impact of the KK modes on the
coefficients $C^{(0){\rm eff}}_{7\gamma}(\mu_b)$ and 
$C^{(0){\rm eff}}_{8G}(\mu_b)$ although this impact is substantially smaller 
than on  $C^{(0)}_{7\gamma}(\mu_W)$ and $C^{(0)}_{8G}(\mu_W)$ in 
(\ref{c2}). This can be understood by investigating the size 
of the different terms in (\ref{C7eff}) and (\ref{C8eff}). Setting 
$\mt = 167\gev$, $1/R=300\gev$, $\mu_b = 5\gev$ and $\as^{(5)}(\mz)
=0.118$ we find 
\begin{eqnarray}
C^{(0){\rm eff}}_{7\gamma}(\mu_b) &=&
0.695 \; C^{(0)}_{7\gamma}(\mu_W) +
0.085 \; C^{(0)}_{8G}(\mu_W) - 0.158 \; C^{(0)}_2(\mu_W)
\nn\\
 &=& 0.695 \; (-0.121) + 0.085 \; (-0.033) - 0.158 = -0.245 \, 
\label{eq:C7geffnum}
\end{eqnarray}
to be compared with $-0.300$ in the SM.
In the absence of QCD corrections we would have 
$C^{(0){\rm eff}}_{7\gamma}(\mu_b) =
C^{(0)}_{7\gamma}(\mu_W)$ (in that case one has $\eta = 1$) and the KK 
effects would be very large as found in section 3. However, 
the additive QCD correction 
present in the last term in (\ref{eq:C7geffnum}) that causes the
large QCD enhancement of the $B\to X_s\gamma$ \cite{Bert,Desh}
screens considerably the effects of the KK modes.
Similar comments apply to $C^{(0){\rm eff}}_{8G}$ for which we find
\begin{eqnarray}
C^{(0){\rm eff}}_{8G}(\mu_b) &=&
0.727 \; C^{(0)}_{8G}(\mu_W) - 0.074 \; C^{(0)}_2(\mu_W)
\nn \\
 &=& 0.727 \; (-0.033) - 0.074 = -0.098 \, 
\label{eq:C8Geffnum}
\end{eqnarray}
to be compared with $-0.144$ in the SM.

We also observe that the strong $\mu_b$-dependence of both coefficients   
\cite{AG1,BMMP:94} is large. 
This scale-uncertainty in the leading order 
can be substantially reduced by including NLO corrections.
\subsection{The Branching Ratio}
In the leading logarithmic approximation one has 
\begin{equation}\label{main}
\frac{Br(B \to X_s \gamma)}
     {Br(B \to X_c e \bar{\nu}_e)}=
 \frac{|V_{ts}^* V_{tb}^{}|^2}{|V_{cb}|^2} 
\frac{6 \alpha}{\pi f(z)} |C^{(0){\rm eff}}_{7}(\mu_b)|^2\,,
\end{equation}
where $f(z)$ with $z=m_c/m_b$ 
is the phase space factor in $Br(B \to X_c e \bar{\nu}_e)$.

The corresponding NLO formulae that include also higher order 
electroweak effects \cite{GAHA} are very complicated and can be found in 
\cite{Gambino:2001ew}.
As reviewed in \cite{AJBMIS,ALIMIS,HURTH}, many groups contributed to 
obtain these NLO results.
In our numerical NLO analysis we benefited enormously from the computer 
programs of the authors of \cite{Gambino:2001ew,GAHA}.

In table~\ref{tab:br} we show the results for $Br(B\to X_s\gamma)$ at LO 
and NLO for the central values of the input parameters in
\cite{Gambino:2001ew} and 
different values of $\mu_b$. While in the LO the $\mu_b$ dependence is 
very sizable, it is very small after including NLO corrections. 
The strong enhancement of the branching ratio by additive QCD corrections 
calculated first in \cite{GRHUWY} and confirmed in \cite{Buras:2002tp} is 
particularly visible at low $1/R$ for which the Wilson coefficients of the 
magnetic operators are strongly suppressed by the KK modes and the additive 
QCD corrections become even more important than in the SM. This table 
demonstrates very clearly the importance of higher order QCD calculations 
in the search for new physics. Without them any comparison between the SM, 
the ACD model and the experimental data would be meaningless.

\begin{figure}[]
\renewcommand{\thesubfigure}{\space(\alph{subfigure})}
\centering
\psfragscanon

  \psfrag{bsgammabsgammabsg}{ $Br(B\rightarrow X_s \gamma )\times 10^{4}$}
  \psfrag{rinvrinv}[][]{ \shortstack{\\  $R^{-1}$ [GeV] }}
  \label{bsg.eps}
        \includegraphics[scale=1]{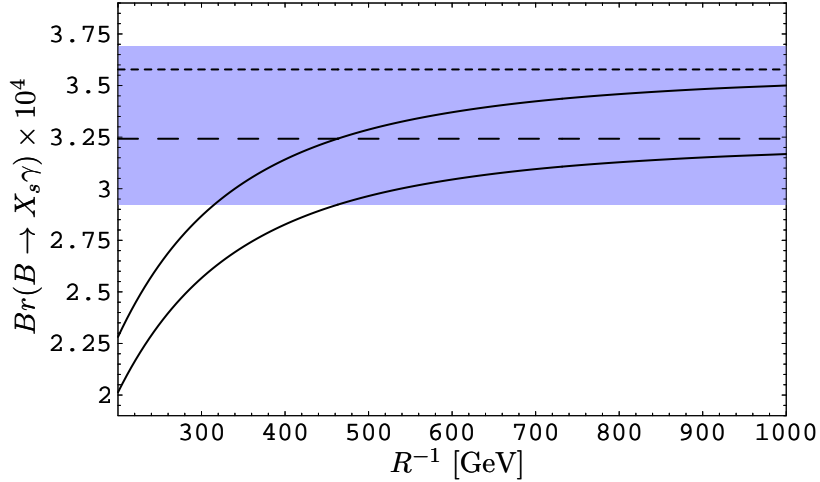}
    \caption[]{\small\label{bsgplot} The branching ratio for $B\to
      X_s\gamma$  and $E_\gamma > 1.6$ GeV as a function of $1/R$. 
See text for
the meaning of various curves.}
  \end{figure}

The strong 
suppression of the branching ratio by the KK modes is clearly seen. 
In fig.~\ref{bsgplot} we compare $Br(B\to X_s\gamma)$ 
in the ACD model with the experimental data and with the expectations 
of the SM. The shaded region represents the data in (\ref{bsgexp}) and the 
upper (lower) dashed horizontal line are the central values in the SM 
for $m_c/m_b=0.22~(m_c/m_b=0.29)$. The solid lines represent the 
corresponding central values in the ACD model. The theoretical errors, 
not shown in the plot, are for all curves roughly $\pm 10\%$.

We observe that in view of the sizable experimental error and
considerable parametric uncertainties in  
the theoretical prediction, the strong suppression of $Br(B\to X_s\gamma)$ 
by the KK modes does not yet provide a powerful lower bound on $1/R$ and the
values $1/R\ge 250\gev$ are fully consistent with the experimental result. 
It should also be emphasized that $Br(B\to X_s\gamma)$ depends
sensitively on the ratio $m_c/m_b$ that 
in (\ref{bsgth}) and table~\ref{tab:br} 
has been set to $0.22$ in accordance 
with the arguments put forward in \cite{Gambino:2001ew}. As seen in 
fig.~\ref{bsgplot}, for 
a value $m_c/m_b=0.29$ that has been used in the past, the branching ratio 
is smaller by roughly $10\%$ and the lower bound on $1/R$ is shifted above 
$400\gev$ if other uncertainties are neglected. 
In order to reduce the dependence on $m_c/m_b$ a NNLO calculation 
is required. Once it is completed and the experimental uncertainties reduced, 
$Br(B\to X_s\gamma)$ may provide a very powerful bound on $1/R$ that is 
substantially stronger than the bounds obtained from the electroweak precision 
data.

The suppression of $Br(B\to X_s\gamma)$ in the ACD model has already been 
found in \cite{AGDEWU}. Our result presented above
is consistent with the one obtained by these authors but differs in details 
as only the dominant diagrams have been taken into account in the latter paper 
and the analysis was performed in the LO approximation.

\begin{table}[htb]
\begin{center}
\begin{tabular}{|c||c|c||c|c||c|c|}
\hline
& \multicolumn{2}{c||}{$\mu_b = 2.5\gev$} &
  \multicolumn{2}{c||}{$\mu_b = 5.0\gev$} &
  \multicolumn{2}{c| }{$\mu_b = 7.5\gev$} \\
\hline
$1/R [\gev]$ & 
Br(LO) & Br(NLO) &
Br(LO) & Br(NLO) &
Br(LO) & Br(NLO) \\
\hline
200 &$ 1.54 $&$ 2.32 $&$ 1.02 $&$ 2.30 $&$ 0.79 $&$ 2.28 $\\
250 &$ 1.92 $&$ 2.66 $&$ 1.37 $&$ 2.65 $&$ 1.11 $&$ 2.63 $\\
300 &$ 2.18 $&$ 2.89 $&$ 1.61 $&$ 2.88 $&$ 1.35 $&$ 2.86 $\\
400 &$ 2.49 $&$ 3.15 $&$ 1.90 $&$ 3.15 $&$ 1.63 $&$ 3.13 $\\
SM  &$ 2.99 $&$ 3.57 $&$ 2.39 $&$ 3.58 $&$ 2.11 $&$ 3.56 $\\
\hline
\end{tabular}
\end{center}
\caption[]{The branching ratio for $B\to X_s\gamma$ in LO and NLO 
in units of $10^{-4}$ as a function of $1/R$ for $m_c/m_b=0.22$ and 
various values of $\mu_b$.
\label{tab:br}}
\end{table}

\section{\boldmath{$B\to X_s~{\rm gluon}$}} 
\setcounter{equation}{0}
\subsection{Preliminaries}
The decay $b\to sg$ is governed by the operator $Q_{8G}$ and consequently the
 value of the coefficient $C_{8G}$ is crucial here. In the SM the branching 
ratio for $b\to sg$ is strongly enhanced by NLO QCD corrections \cite{GL00}
 with the result
\be\label{bsg1}
Br(b\to sg)=(5.0\pm 1.0)\cdot 10^{-3}, \qquad m_c/m_b=0.29
\ee
which is larger by a factor of 2.5 than the LO value. The sizable uncertainty
shown in (\ref{bsg1}) is due solely to the renormalization scale dependence. 
The other important uncertainty is the value of $m_c/m_b$. For $m_c/m_b=0.22$
we find using the computer program of \cite{GL00}
\be\label{bsg2}
Br(b\to sg)=(4.1\pm 0.7)\cdot 10^{-3}, \qquad m_c/m_b=0.22~.
\ee

\begin{figure}[]\label{BSGLUE}
\renewcommand{\thesubfigure}{\space(\alph{subfigure})}
\centering
\psfragscanon

  \psfrag{bsgluonbsgluon}{ $Br(B\rightarrow X_s g )\times 10^3$}
  \psfrag{rinvrinv}[][]{  \shortstack{\\  $R^{-1}$ [GeV]} }
  \label{bsgluon.eps}
        \includegraphics[scale=1]{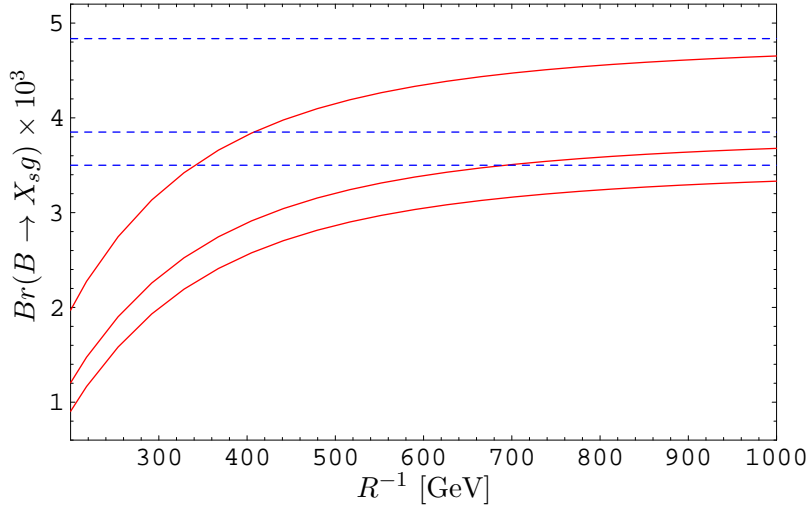}
    \caption[]{\small\label{bsgluonplot} The branching ratio for $B\to
      X_s g$ for $m_c/m_b = 0.22$ and $\mu_b$ = $2.5$, $5$, $7.5$ GeV. 
$Br(B\rightarrow X_s g )$ decreases with increasing $\mu_b$.
The 
dashed lines represent the SM prediction.} 
  \end{figure}

In fig.~\ref{bsgluonplot} we show the results for $Br(b\to sg)$ in the
ACD model as a function 
of $1/R$ for three values of $\mu_b$ and $m_c/m_b=0.22$. In obtaining these 
results we benefited enormously from the computer 
program of the authors of \cite{GL00}. As anticipated in section 3, 
there is a very strong suppression of the branching ratio in question by 
the KK contributions and if the strong $\mu_b$ and $m_c/m_b$ dependences
could be put under control and the branching ratio could somehow be extracted
from the data, the decay $B\to X_s~{\rm gluon}$ could offer a very powerful 
constraint on $1/R$. Indeed, even for $1/R=600\gev$ a clear distinction between 
the SM and the ACD model predictions can be made for a fixed $\mu_b$.

Unfortunately, it will take some time before 
the strong $\mu_b$ and $m_c/m_b$ dependences
can be significantly reduced and the extraction of the branching ratio from 
the experimental data is very difficult if not
impossible~\cite{GL00,Lenz:1997aa}. 
Yet, it is not 
excluded that one day this result could become relevant.  

\section{$B\to X_{\lowercase{s}} \lowercase{\mu}^+\lowercase{\mu}^-$}
         \label{sec:Heff:BXsee:nlo}
\setcounter{equation}{0}
\subsection{Preliminaries}
The inclusive $B\to X_s\mu^+\mu^-$ decay has been the subject of very 
intensive theoretical and experimental studies during the last 15 years. On 
the experimental side only the BELLE collaboration reported the
observation of this decay with \cite{Belle02}
\be\label{bsxmmexp}
Br(B\to X_s\mu^+\mu^-)= (7.9\pm 2.1^{+2.0}_{-1.5})\cdot 10^{-6}~.
\ee 
For the decay to be
dominated by perturbative contributions one has to remove $\bar c c$
resonances by appropriate kinematical cuts in the dilepton mass
spectrum.
The SM expectation \cite{ALIHILL} for the low dilepton mass window is given by
\be\label{bsxmmth}
\tilde Br(B\to X_s\mu^+\mu^-)_{\rm SM}= (2.75\pm 0.45)\cdot 10^{-6}~,
\ee
where the dilepton mass spectrum has been integrated in the limits
suggested by BELLE~\cite{Belle02} and given in
(\ref{intlimits}). 

This cannot be directly compared to the
experimental result in (\ref{bsxmmexp}) that is supposed to include the
contributions from the full dilepton mass spectrum. 
Fortunately future experimental
analyses are supposed to give  the results corresponding to the low dilepton 
mass window so that a direct comparison between experiment and  
theory will be possible.

The most recent reviews summarizing the theoretical status can be found 
in \cite{HURTH,ALIHILL}. The purpose of this section is  the calculation of
the impact of the KK modes on the branching ratio, the invariant dilepton 
mass spectrum and the forward-backward asymmetry.  

In contrast to the $B \to X_s \gamma$ decay the inclusion of NLO corrections 
to $Br(B \to X_s \mu^+ \mu^-)$ does not require the calculation of QCD 
corrections to the relevant penguin and box diagrams. 
Detailed explanation of this feature is given in \cite{AJBLH,HURTH,BBL}. 
Consequently in this
decay we are in the position to present the complete NLO analysis  
in the ACD model. On the other hand the NNLO corrections to 
$B\to X_s\mu^+ \mu^-$ 
have been calculated by now within the SM \cite{Bobeth,NNLO1,NNLO2}
and it will be of interest to include these corrections also in the 
ACD model following the philosophy as in the case of NLO corrections 
to $B\to X_s\gamma$ and $B\to X_s~{\rm gluon}$ decays. However, in order to
see the impact of the KK modes on $B\to X_s\mu^+ \mu^-$ in a transparent 
manner we will first present a NLO analysis of this decay.

\subsection{Effective Hamiltonian} 
The effective Hamiltonian  at scales $\mu=O(m_b)$ is
 given by
\begin{equation} \label{Heff2_at_mu}
{\cal H}_{\rm eff}(b\to s \mu^+\mu^-) =
{\cal H}_{\rm eff}(b\to s\gamma)  - \frac{G_{\rm F}}{\sqrt{2}} V_{ts}^* V_{tb}
\left[ C_{9V}(\mu) Q_{9V}+
C_{10A}(M_W) Q_{10A}    \right]\,,
\end{equation}
where we have again neglected the term proportional to $V_{us}^*V_{ub}$
and ${\cal H}_{\rm eff}(b\to s\gamma)$ is given in (\ref{Heff_at_mu}).
In addition to the operators relevant for $B\to X_s\gamma$,
there are two new operators:
\begin{equation}\label{Q9V}
Q_{9V}    = (\bar{s} b)_{V-A}  (\bar{\mu}\mu)_V\,,         
\qquad
Q_{10A}  =  (\bar{s} b)_{V-A}  (\bar{\mu}\mu)_A\,,
\end{equation}
where $V$ and $A$ refer to $\gamma_{\mu}$ and $ \gamma_{\mu}\gamma_5$,
respectively. They are generated through the electroweak
penguin diagrams of fig.~\ref{penguindiagrams} 
 and the related box diagrams needed mainly
to keep gauge invariance.

It is convenient to work with the coefficients $\tilde C_{9}$ and
$\tilde C_{10}$ defined by 
\begin{equation} \label{C10}
C_{9V}(\mu) = \frac{\alpha}{2\pi} \tilde C_9(\mu), \qquad
C_{10A}(\mu) =  \frac{\alpha}{2\pi} \tilde C_{10}(\mu).
\ee
Actually $Q_{10A}$ does not renormalize
under QCD and  its coefficient is $\mu$ independent with
\be
\tilde C_{10}(\mu) = - \frac{Y(x_t,1/R)}{\sin^2\theta_{w}}
\end{equation}
and $Y(x_t,1/R)$ defined in (\ref{yyx}) and calculated in \cite{BSW02}.

The NLO QCD corrections to $\tilde C_{9}(\mu)$ in the SM model 
have been calculated in 
\cite{Mis:94,BuMu:94}. Generalizing this result to the ACD model 
we obtain  in the NDR scheme
\begin{equation}\label{C9tilde}
\Ctilde_9^{\rm NDR}(\mu)  =  
P_0^{\rm NDR} + \frac{Y(x_t,1/R)}{\sin^2\theta_{w}} -4 Z(x_t,1/R) +
P_E E(x_t,1/R)
\end{equation}
with
\be
\label{P0NDR}
P_0^{\rm NDR} = 2.60\pm 0.25
\ee
as in the SM. The uncertainty comes from the $\mu_b$ dependence, with 
the highest value corresponding to $\mu_b=2.5\gev$ and the lowest to 
$\mu_b=7.5\gev$. The $\mu_b$ dependence is substantially reduced through 
the $\mu_b$ dependence of the perturbatively calculable matrix elements 
of the operators involved.
Analytic formula for $P_0^{\rm NDR}$ can be found in 
\cite{Mis:94,BuMu:94}.
The functions $Y$, $Z$ and $E$ are given in section 3.
$P_E$ is ${\cal O}(10^{-2})$ and
consequently the last term in (\ref{C9tilde}) can be neglected. 

In table~\ref{tab:C9} we
show the $\Ctilde_9^{\rm NDR}(\mu_b)$ as a function of  $1/R$ 
for  $\mt= 167\gev$, different values of $\mu_b$ and $\alpha_s(M_Z)=0.118$.
We observe that the impact of the KK modes on this Wilson coefficient is very
small due to the approximate cancellation of these contributions to $Y$ and
$Z$. Consequently the impact of the KK modes on this decay proceeds almost
entirely through the enhancement of the coefficient $\tilde C_{10}$ and the 
suppression of $C^{(0){\rm eff}}_{7\gamma}$.

\begin{table}[htb]
\begin{center}
\begin{tabular}{|c||c|c|c|}
\hline
$1/R [\gev]$ & $\mu_b = 2.5\gev$ &
  $\mu_b = 5.0\gev$ &
$\mu_b = 7.5\gev$\\
\hline
200 & 4.476  & 4.226  & 3.976  \\
250 & 4.434  & 4.184 & 3.934 \\
300 & 4.413  & 4.163 & 3.913  \\
400 & 4.393  & 4.143  & 3.893  \\
SM &  4.372   & 4.122 & 3.872 \\
\hline
\end{tabular}
\end{center}

\caption[]{The values of $\Ctilde_9^{\rm NDR}(\mu_b)$ in NLO
for $\mt = 167 \gev$ as a function of $1/R$ and 
various values of $\mu_b$.
\label{tab:C9}}
\end{table}

\subsection{The Differential Decay Rate}
         \label{sec:Heff:BXsee:nlo:rate}
We are now ready to present the results for the differential decay
rate based on the effective Hamiltonian in (\ref{Heff2_at_mu}) and
the spectator model for the matrix elements of $Q_i$.

Introducing
\begin{equation} \label{invleptmass}
\hat s = \frac{(p_{\mu^+} + p_{\mu^-})^2}{\mb^2}, \qquad z =
\frac{\mc}{\mb}
\end{equation}
and calculating the one-loop matrix elements of $Q_i$ using the
spectator model in the NDR scheme one finds \cite{Mis:94,BuMu:94}
\be \label{rateee}
\tilde T(\hat s) \equiv \frac{{d}/{d\hat s} \, 
\Gamma (b \to s \mu^+\mu^-)}{\Gamma
(b \to c e\bar\nu)} = \frac{\alpha^2}{4\pi^2}
\left|\frac{V_{ts}}{V_{cb}}\right|^2 \frac{(1-\hat s)^2}{f(z)\kappa(z)}
U(\hat s)
\ee
where
\be\label{US} 
U(\hat s)=
(1+2\hat s)\left(|\Ctilde_9^{\rm eff}(\hat s)|^2 + |\Ctilde_{10}|^2\right) + 
4 \left( 1 + \frac{2}{\hat s}\right) |C_{7\gamma}^{(0){\rm eff}}|^2 + 12
C_{7\gamma}^{(0){\rm eff}} \ \RE\,\Ctilde_9^{\rm eff}(\hat s) 
\ee
and $\Ctilde_9^{\rm eff}(\hat s)$ is a function of $\hat s$ that depends on
the Wilson coefficient $\Ctilde_9^{\rm NDR}$ and includes also contributions 
from four quark operators. Explicit formula can be found in
\cite{Mis:94,BuMu:94}.

Next,  $f(z)$ is the phase-space factor for $B\to X_c
e\bar\nu$  and $\kappa(z)=0.88$ 
\cite{CM78,KIMM} is the
corresponding QCD correction. 
In our numerical calculations we set $Br(B\to X_c e\bar\nu)_{\rm exp}=0.104$.
Finally, the dilepton mass spectrum is defined as
\be
   \frac{d}{d \hat s} Br(B \to X_s \mu^+ \mu^-) (\hat s) =  Br(B\to X_c
   e\bar\nu)_{\rm exp} \times \tilde T(\hat s) \equiv T(\hat s)~. 
\ee

Before presenting the numerical analysis let us note that among the
different terms in (\ref{US}), $|\Ctilde_9^{\rm eff}(\hat s)|^2$ is 
essentially as in the SM, $|\Ctilde_{10}|^2$ is significantly 
enhanced, while the last two terms in (\ref{US}) are suppressed. 
However, the last term in (\ref{US}) is negative and consequently its
suppression results in an enhancement of $U(\hat s)$ in addition to the 
one due to $\Ctilde_{10}$. 

The formulae presented above served mainly to illustrate the pattern of 
various contributions that is not modified significantly at the NNLO 
level \cite{Bobeth,NNLO1,NNLO2}. However,  
the latter corrections cannot be neglected and we have 
included them in our analysis.
In obtaining these results we benefited enormously from the computer 
program of the authors of \cite{ALIHILL,NNLO1} that we generalized to
the ACD model.

\begin{figure}[hbt]
\renewcommand{\thesubfigure}{\space(\alph{subfigure})} 
  \centering 
\subfigure[]{\psfragscanon
  \psfrag{brsllbrsllbrsllbrsll}{ $\frac{d}{d \hat s} Br(B\rightarrow X_s l^+ l^- )\times 10^5$}
  \psfrag{hats}[][]{ \shortstack{\\  $\hat{s}$  }}
    \resizebox{.36\paperwidth}{!}{\includegraphics[]{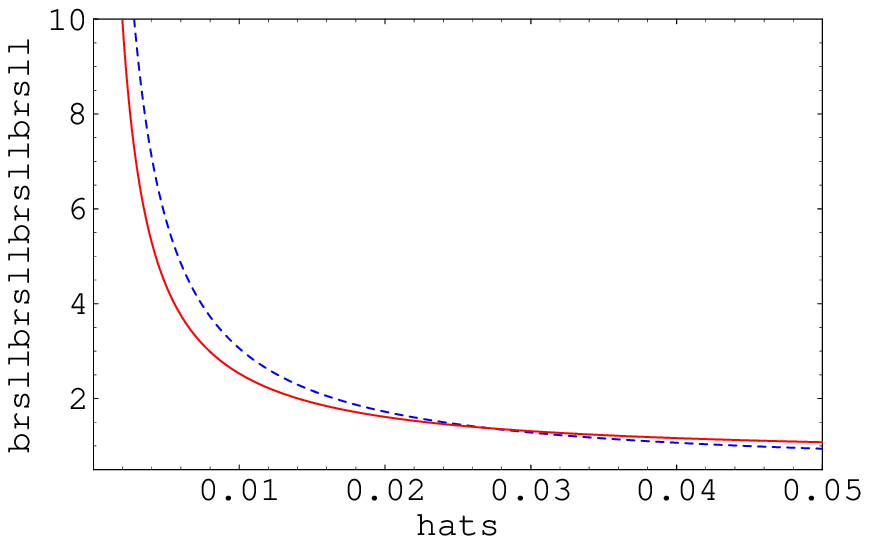}}
    }
  \subfigure[]{\psfragscanon
  \psfrag{brsllbrsllbrsllbrsll}{  $\frac{d}{d \hat s}
  Br(B\rightarrow X_s l^+ l^- )\times 10^5$}
  \psfrag{hats}[][]{ \shortstack{\\  $\hat{s}$ }}
    \resizebox{.36\paperwidth}{!}{
    \includegraphics[]{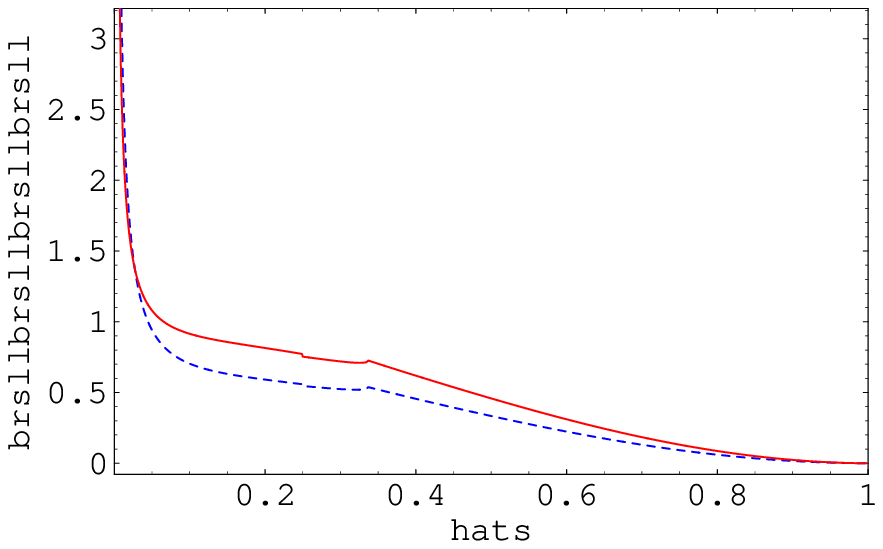}}
    }

  \caption[]{\small\label{brsll} $\frac{d}{d \hat s} Br(B\rightarrow X_s l^+
  l^- )$  in the SM (dashed line) and the ACD model for $R^{-1} =250$
  GeV. (a) Low $\hat s$ region and (b) full dilepton mass spectrum.}
\end{figure}

In fig.~\ref{brsll} we show $T(\hat s)$ as a function of $\hat s$ for 
 $1/R=250\gev$. As anticipated before, the KK contributions enhance  
$T(\hat s)$ with respect to the SM, see
  fig.~\ref{brsll} (b), except for very small $\hat s$ where 
the contribution of the coefficient $C_{7\gamma}^{(0){\rm eff}}$ dominates 
and its suppression results in the suppression of $T(\hat s)$, 
see fig.~\ref{brsll} (a). 
\begin{figure}[hbt]
\renewcommand{\thesubfigure}{\space(\alph{subfigure})} 
  \centering 
\psfragscanon
  \psfrag{brsllbrsllbrsll}{  $\tilde Br(B\rightarrow X_s \mu^+ \mu^- )\times 10^{6}$}
  \psfrag{rinvrinv}[][]{ \shortstack{\\ $R^{-1}$ [GeV] }}
    \resizebox{.5\paperwidth}{!}{
    \includegraphics[]{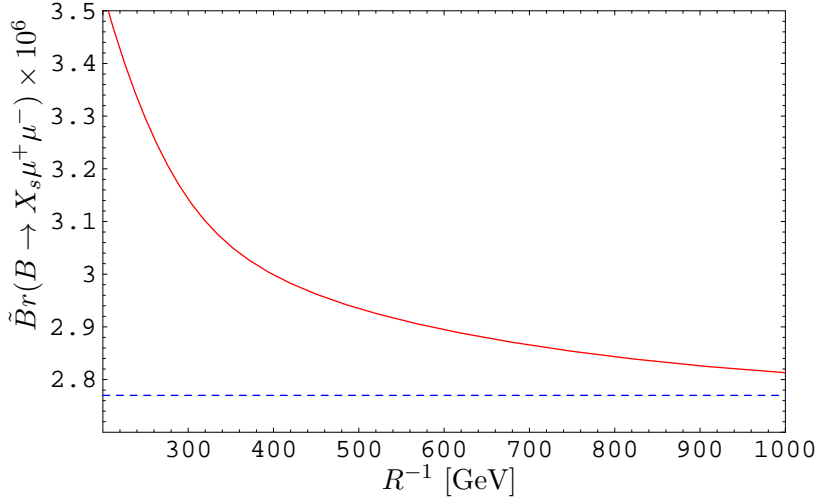}}
    
  \caption[]{\small\label{brsllRinv} $\tilde Br(B\rightarrow X_s \mu^+
  \mu^- )$  in the SM (dashed line) and in the ACD model. The
  integration limits have been chosen as defined in (\ref{intlimits}).}
\end{figure}

At this point the following remarks are in order. The theoretical
calculations are cleanest for $\hat s_0\le 0.25$ where the NNLO
calculations for the inclusive decays have been completed
\cite{NNLO1,NNLO2} and resonant  
effects due to $J/\psi$, $\psi^\prime$, etc. are expected to be small. For 
this reason as done in \cite{ALIHILL}, we will first present the branching 
ratio for $B\to X_s\mu^+\mu^-$ by integrating over $\hat s$ only in the 
region
\be\label{intlimits}
\left(\frac{2 m_\mu}{m_b}\right)^2\le\hat s\le
\left(\frac{M_{J/\psi}-0.35\gev}{m_b}\right)^2
\ee
that gives  in the case of the SM \cite{ALIHILL} the result in
(\ref{bsxmmth}) 
where the error comes from the variation of $\mu_b$,
$m_t^{\rm pole}$ and $m_c/m_b$.

In fig.~\ref{brsllRinv} we show the branching ratio 
$\tilde Br(B\to X_s\mu^+\mu^-)$ as a function
of $1/R$. In obtaining these results we followed closely the procedure 
of the authors of \cite{ALIHILL,NNLO1} and generalized their computer
programs  
to include the KK contributions. We observe a modest enhancement 
of  $\tilde Br(B\to X_s\mu^+\mu^-)$ that for $1/R=300\gev$ amounts to 
roughly $12\%$.

Now the experimental number in (\ref{bsxmmexp}) refers to the so-called 
non-resonant branching ratio integrated over the entire dilepton invariant 
mass spectrum. As we do not have any access to the experimental analysis and 
the way the resonance contributions have been removed, we will follow the 
authors of \cite{ALIHILL} who integrated the theoretical expressions over 
the whole dilepton invariant mass spectrum and obtained in the SM the result
$Br(B\to X_s\mu^+\mu^-)_{\rm SM} = (4.15\pm 0.7)\cdot 10^{-6}~$. 
 For instance for $1/R=300\gev$ we find 
\be\label{bsxmmACD}
Br(B\to X_s\mu^+\mu^-)_{\rm ACD}= (4.8 \pm 0.8)\cdot 10^{-6}~.
\ee

The enhancement of the relevant branching ratio is now 
stronger than in the previous case as the  role of $\tilde C_{10}$ becomes 
more important for larger $\hat s$.
The result in (\ref{bsxmmACD}) is significantly closer to Belle's result 
in (\ref{bsxmmexp}) than 
the SM prediction. However, the very large experimental error and 
still sizable 
theoretical uncertainties  in the branching ratio corresponding to the full 
dilepton mass spectrum, also of non-perturbative origin,
 preclude definite conclusions at present. As we stated before it is safer to
consider the branching ratio for the low dilepton mass window as given in 
in fig.~\ref{brsllRinv}.

\subsection{Forward-Backward Asymmetry}

Of particular interest is the Forward-Backward asymmetry in 
$B\to X_s\mu^+\mu^-$. It becomes non-zero only at the NLO level and 
is given in this approximation by 
\cite{AMM}
\be\label{ABF}
A_{\rm FB}(\hat s)=\frac{1}{\Gamma
(b \to c e\bar\nu)} \int_{-1}^1 d \cos \theta_l \frac{ d^2
  \Gamma (b \to s \mu^+\mu^-)} { d\hat s d\cos \theta_l} {\rm sgn}
(\cos \theta_l) = -3 \tilde C_{10}
\frac{\left[\hat s \RE\,\Ctilde_9^{\rm eff}(\hat s)
+2 C_{7\gamma}^{(0){\rm eff}}\right]}
{U(\hat s)}
\ee
with $U(\hat s)$ given in (\ref{US}) and where $\theta_l$ is the angle
between the $\mu^+$ and $B$ meson momenta in the center of mass
frame. Similar to the case of exclusive 
decays \cite{Burdman}, the asymmetry $A_{\rm FB}(\hat s)$ vanishes at 
$\hat s=\hat s_0$ that in the case of the inclusive decay considered 
is determined through 
\be\label{ZERO}
\hat s_0 \RE\,\Ctilde_9^{\rm eff}(\hat s_0)+2 C_{7\gamma}^{(0){\rm eff}}=0.
\ee
The fact that $A_{\rm FB}(\hat s)$ and the value of $\hat s_0$,
being sensitive to short distance physics, are in addition 
subject to only very small non-perturbative uncertainties makes them 
particularly useful quantities to test for physics beyond the SM. 

\begin{figure}[hbt]
\renewcommand{\thesubfigure}{\space(\alph{subfigure})} 
  \centering 
  \subfigure[]{\psfragscanon
  \psfrag{nfbnfb}{ $\hat{A}_{FB}$}
  \psfrag{hats}[][]{ \shortstack{\\ $\hat{s} $ }}
      \resizebox{.36\paperwidth}{!}{\includegraphics[]{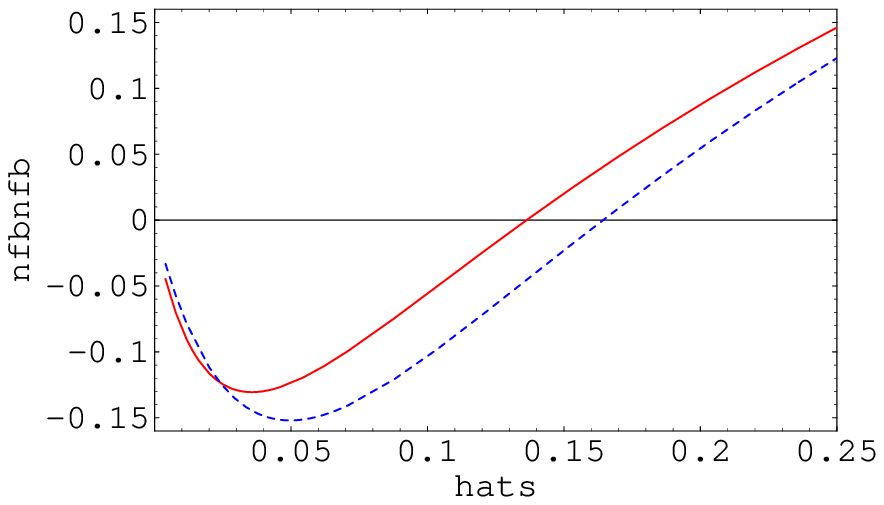}}
    }
  \subfigure[]{\psfragscanon
  \psfrag{zafb}{ $\hat{s}_{0}$}
  \psfrag{rinvrinv}[][]{ \shortstack{\\  $R^{-1}$ [GeV]  }}
    \resizebox{.36\paperwidth}{!}{ \includegraphics[]{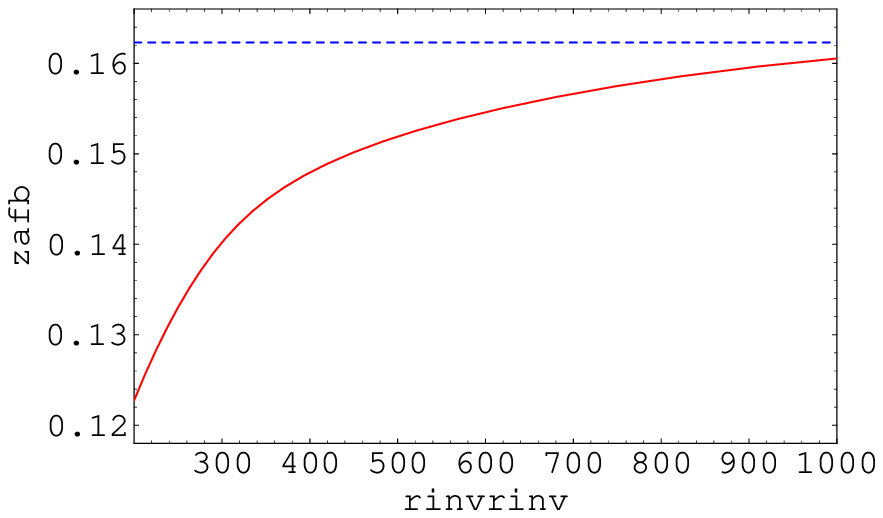}}
    }
  \caption[]{\small\label{normalizedfb} (a) Normalized Forward-Backward
    asymmetry in the SM (dashed line) and ACD for $R^{-1}=250$ GeV. 
(b) Zero of the forward backward
    asymmetry $A_{FB}$ in the SM (dashed line) and the ACD model.}
\end{figure}

The formulae (\ref{ABF}) and (\ref{ZERO}) have recently been
generalized to include NNLO corrections \cite{NNLO1,NNLO2} that turn out to 
be significant. 
In particular they shift the NLO value of $\hat s_0$ from $0.14\pm
0.02$ to $0.162 \pm 0.008$. 
In fig.~\ref{normalizedfb} (a)  we show the normalized
Forward-Backward asymmetry\footnote{The normalized FB asymmetry is
  given as $\hat{A}_{\rm FB}(\hat s) =\Gamma
(b \to c e\bar\nu) \times A_{\rm FB}(\hat s)/ \int_{-1}^1 d \cos \theta_l \frac{ d^2
  \Gamma (b \to s \mu^+\mu^-)} { d\hat s d\cos \theta_l}$ }
$\hat{A}_{\rm FB}(\hat s)$  for the central values of the input parameters
 that we obtained by means of the formulae and the computer program of 
\cite{ALIHILL,NNLO1} modified by the 
KK contributions calculated here. The dependence of $\hat s_0$ on $1/R$  
is shown in fig.~\ref{normalizedfb} (b).

We observe that the value of $\hat s_0$ is considerably reduced relative 
to the SM result obtained including NNLO corrections 
\cite{ALIHILL,NNLO1,NNLO2}. This decrease originates 
in the suppression of the coefficient $C_{7\gamma}^{(0){\rm eff}}$ as 
clearly seen in (\ref{ZERO}). We recall that $\Ctilde_9^{\rm eff}$ is 
only weakly affected by the KK contributions.  
For 
$1/R=300\gev$ we find a value for $\hat s_0$ that is  very close to 
the NLO prediction of the 
SM. This result demonstrates very clearly the importance of the higher 
order QCD corrections, in particular in quantities like $\hat s_0$ 
that are theoretically clean. We expect that the results in
figs.~\ref{normalizedfb} (a) and (b) 
will play an important role in the tests of the ACD model in the future.

\subsection{Correlation between \boldmath{$\hat s_0$} and 
\boldmath{$Br(B\to X_s\gamma)$}}

In MFV models there exist a number of correlations between different 
measurable 
quantities that do not depend on specific parameters of a given model. These 
correlations originate in the general property of these models: the  
new physics enters only through the Inami-Lim (IL) functions. 
As several processes depend on the same IL functions, eliminating these 
functions in favour of measurable quantities allows to derive interesting 
universal relations between these quantities that do not involve parameters 
specific to a given model. Examples can be found in \cite{UUT,REL}.

Here we would like to point out a correlation between $\hat s_0$ and 
$Br(B\to X_s\gamma)$ that is present in the ACD model and in any MFV model in
which new physics contributions to $\Ctilde_9^{\rm eff}$ and its $\hat s$ 
dependence  can be neglected to 
first approximation. This is the case of the ACD model and of a large 
class of supersymmetric models discussed for instance in \cite{ALIHILL}. 
As new physics contributions to 
the $Z^0$-penguin represented by the function $C$ are suppressed in 
$\Ctilde_9^{\rm eff}$ by the factor $(4\sin^2\theta_w-1)$, 
the necessary requirement for the weak dependence of $\Ctilde_9^{\rm eff}$ on 
new physics is the smallness of these contributions to the $B$ and $D$
functions. This in fact is the case for many MFV models.

If $\Ctilde_9^{\rm eff}$ is unaffected by new physics contributions and its 
$\hat s$--dependence is weak then, as seen in (\ref{ZERO}), $\hat s_0$
is simply  
proportional to $C_{7\gamma}^{(0){\rm eff}}$ with the proportionality factor 
independent of new physics.
\begin{figure}[hbt]

  \centering 
 \psfragscanon
  \psfrag{bsgammabsgammabsgamma}{ \shortstack{\\ \\ $(Br(B\to
 X_s\gamma)\times 10^4)^\frac12$ ${}$}}
  \psfrag{hats0}[][]{ \shortstack{\\ $\hat{s}_0 $ }}
      \resizebox{.36\paperwidth}{!}{\includegraphics[]{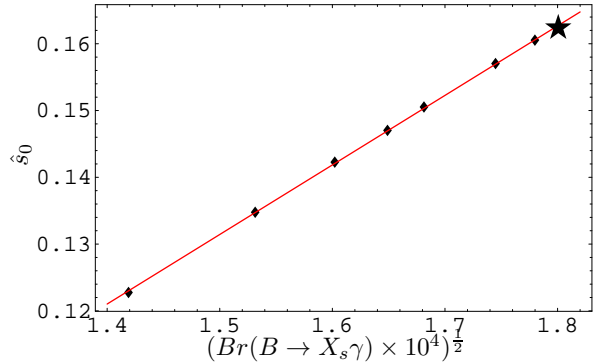}}
    
  \caption[]{\small\label{corrplot} Correlation between
    $\sqrt{Br(B\to X_s\gamma)}$  and $\hat s_0$. The straight line is
    a least square fit to a linear function. The dots are the results in the ACD
    model
    for $1/R = 200,250,300,350,400,600$ and $1000$ GeV  and the star
    denotes the SM value.
}
\end{figure}
Consequently $\hat s_0$ is proportional to $\sqrt{Br(B\to X_s\gamma)}$,
\be
\hat s_0 \propto \frac{\sqrt{Br(B\to X_s\gamma)}}{\Ctilde_9^{\rm eff}},
\ee
and  can simply be predicted from the 
measured value of $Br(B\to X_s\gamma)$  and $\Ctilde_9^{\rm eff}$ given in 
the SM. 
This correlation is shown in fig.~\ref{corrplot}. It depends of course on the relation
between $C_{7\gamma}^{(0){\rm eff}}$  and $Br(B\to X_s\gamma)$, in particular
on $m_c/m_b$. In fig.~\ref{corrplot} we have used $m_c/m_b=0.29$. 
The plot in fig.~\ref{corrplot} has been obtained within the ACD model by expressing 
$1/R$  in fig.~\ref{normalizedfb} (b)   through $Br(B\to X_s\gamma)$ by means
of the results in  fig.~\ref{bsgplot}. 
The fact that the result of this exercise is a straight line confirms
our result that the new physics contributions to $\Ctilde_9^{\rm eff}$ 
and its $\hat s$ dependence are small.

\section{\boldmath{$K_L\to \pi^0 e^+ e^-$}}
\subsection{Preliminaries}
There are three contributions to this decay: CP-conserving,
indirectly CP-violating and directly CP-violating.
Unfortunately out of these three contributions only the
directly CP-violating one can be calculated reliably.
In this contribution there are
practically no theoretical uncertainties related to hadronic matrix
elements because $\langle\pi^0|(\bar sd)_{V-A}| K_{\rm L}\rangle$ can be
extracted using isospin symmetry from the well measured decay
$K^+\to\pi^0e^+\nu$. In what follows we will only consider this 
contribution. Most recent reviews of $K_L\to\pi^0e^+e^-$ with references
to earlier literature can be found in \cite{Isidori}. 

\subsection{The Branching Ratio}
Generalizing the NLO formula in \cite{BLMM} to the ACD model we obtain 
\begin{equation}\label{9a}
Br(K_{\rm L} \to \pi^0 e^+ e^-)_{\rm dir} = \kappa_e(\IM\lambda_t)^2
(\tilde y_{7A}^2 + \tilde y_{7V}^2)\,,
\end{equation}
where
$\IM \lambda_t = \IM (V_{td} V^*_{ts})$,
\begin{equation}
\label{kappae}
\kappa_e=\frac{1}{V_{us}^2}\frac{\tau(K_{\rm L})}{\tau(K^+)}
\left( \frac{\alpha}{2 \pi} \right)^2 Br(K^+\to\pi^0e^+\nu)
=6.34 \cdot 10^{-6}
\end{equation}
and
\begin{equation}\label{y7vpbe}
\tilde{y}_{7V} =
P_0 + \frac{Y(x_t,1/R)}{\sin^2\theta_{ w}} - 4 Z(x_t,1/R)+ P_E E(x_t,1/R)~,
\end{equation}
\begin{equation}\label{y7apbe}
\tilde{y}_{7A}=-\frac{1}{\sin^2\theta_{w}} Y(x_t,1/R)
\end{equation}
with $Y(x_t,1/R)$, $Z(x_t,1/R)$ and $E(x_t,1/R)$  given in section 3. 
The formula  (\ref{y7vpbe}) has a structure identical to 
(\ref{C9tilde}) relevant for
$B \to X_s \mu^+ \mu^-$ with different numerical values for 
$P_0$ and $P_E$ due to different scales $\mu$ involved. 
$P_E$ is $\ord(10^{-2})$ and consequently the last
term in (\ref{y7vpbe}) can be neglected. 
The next-to-leading QCD corrections to the coefficients above enter
only $P_0$. They
have been calculated in \cite{BLMM}. One finds $P_0=3.05\pm0.06$, where 
the error comes from the $\alpha_s$ and scale uncertainties. 
Similarly to $B \to X_s \mu^+ \mu^-$, the effect of KK contributions is mainly 
felt in $\tilde{y}_{7A}$.

The present experimental bound
 from KTeV \cite{kTev}
\begin{equation}
  \label{eq:brklexp}
  Br(\kpe)<3.5 \cdot 10^{-10}~(90\% C.L.)
\end{equation}
is still by two orders of magnitude away from the theoretical
expectations in the SM and the ACD model. 
\subsection{Numerical Analysis}
In table~\ref{BKL} we show $Br(K_{\rm L} \to \pi^0 e^+ e^-)_{\rm dir}$ in the ACD 
model as a function of $1/R$ for $P_0=3.05$ and $m_t=167\gev$. We also show 
there $\IM\lambda_t$. To this end we have used the analysis of the unitarity 
triangle of \cite{BSW02}.  
The enhancement of $Br(K_{\rm L} \to \pi^0 e^+ e^-)_{\rm dir}$ is at most 
$10\%$ because the enhancement of the function $Y$ in $\tilde y_{7A}$ 
is compensated to a large 
extent by the suppression of $\IM\lambda_t$. 

\begin{table}[hbt]
\begin{center}
\begin{tabular}{|c||c|c|c|c|c|}\hline
 $1/R~[{\rm GeV}]$  & {$200$} & {$250$}& {$300$} & {$400$} & {${\rm SM}$} 
 \\ \hline
$\IM\lambda_t\times 10^4$ & $ 1.202 $ &  $1.250 $ &  $1.276$ & $1.302$ 
&$ 1.333$    
\\ \hline
$Br(K_{\rm L} \to \pi^0 e^+ e^-)_{\rm dir}\times 10^{12}$ 
& $ 4.87 $ &  $4.77 $ &  $ 4.69$ & $4.58$ & $ 4.39 $  
\\ \hline
\end{tabular}
\end{center}
\caption[]{\small Values for $\IM\lambda_t$ and 
$Br(K_{\rm L} \to \pi^0 e^+ e^-)_{\rm dir}$ for different $1/R$.  
\label{BKL}}
\end{table}

\section{The Ratio \boldmath{$\epe$}}
\setcounter{equation}{0}
\subsection{Preliminaries}
The ratio $\epe$ that parametrizes the size of the direct CP violation with 
respect to the indirect CP violation in $K_L\to \pi\pi$ decays has been the 
subject of very intensive experimental and theoretical studies. On the
experimental side the world average based on the recent results from 
NA48 \cite{NA48} and KTeV \cite{KTeV} and previous results from NA31 and 
E731 collaborations reads
\begin{equation}
  \label{eps}
  \epe=(16.6\pm 1.6) \cdot 10^{-4}~.
\end{equation}
On the other hand, the theoretical estimates of this ratio are subject 
to very large hadronic uncertainties. While several analyses within the 
SM find results that are compatible with (\ref{eps}), it is fair to say 
that the chapter on the theoretical calculations of $\epe$ is certainly 
open. Most recent reviews with references to original papers can be found 
in \cite{epsref,Hambye:2003cy,BJ03}. 

In view of this situation, our strategy will be to choose various sets of 
the values of
the main non-perturbative parameters $B_6^{(1/2)}$ and $B_8^{(3/2)}$ for
a fixed strange quark mass $\ms(\mc)$ and to 
investigate for each chosen set ($B_6^{(1/2)},B_8^{(3/2)})$ 
the impact of the KK modes on $\epe$. 

\subsection{Basic Formula}
The
formula for $\epe$ of \cite{EP99,BRMSSM} generalizes to the ACD model
as follows: 
\be \frac{\varepsilon'}{\varepsilon}= \IM\lambda_t
\cdot F_{\varepsilon'}(x_t,1/R)~,
\label{epeth}
\ee
where
\be
F_{\varepsilon'}(x_t,1/R) =P_0 + P_X \, X(x_t,1/R) + 
P_Y \, Y(x_t,1/R) + P_Z \, Z(x_t,1/R)+ P_E \, E(x_t,1/R)
\label{FE}
\ee
with
the functions $X$, $Y$, $Z$, $E$ given in section 3.
Strictly speaking the functions $X$ and $Y$ entering (\ref{FE}) are 
given by $X=C+ B^{(uu)}$ and $Y=C+B^{(dd)}$ and consequently they
differ slightly from the ones defined in (\ref{xx9}) and (\ref{yyx})
that involve $B^{\nu\bar\nu}$ and $B^{\mu\bar\mu}$ instead 
of $B^{(uu)}$ and $B^{(dd)}$, respectively. However, this difference 
caused only by the KK contributions 
to the box diagrams is fully negligible and the numerical values for 
$X$ and $Y$ given in table~\ref{XYZ} can also be used here.

The coefficients $P_i$ are given in terms of the non-perturbative
parameters $B_6^{(1/2)}$ and $B_8^{(3/2)}$ and the strange quark
mass  $\ms(\mc)$ as follows:
\begin{equation}
P_i = r_i^{(0)} + 
r_i^{(6)} R_6 + r_i^{(8)} R_8 \,~,
\label{eq:pbePi}
\end{equation}
where
\be\label{RS}
R_6\equiv \bsi\left[ \frac{121\mev}{\ms(\mc)+\md(\mc)} \right]^2,
\qquad
R_8\equiv \bei\left[ \frac{121\mev}{\ms(\mc)+\md(\mc)} \right]^2.
\ee
$\bsi$ and $\bei$ parameterize the matrix elements of the dominant
QCD-penguin ($Q_6$) and the dominant electroweak penguin ($Q_8$)
operator, respectively.
The numerical values of $r_i^{(0)}$, $r_i^{(6)}$ and $r_i^{(8)}$ 
for different values of $\Lms^{(4)}$ at $\mu=\mc$ have recently been 
updated in \cite{BJ03} and are given in the NDR 
renormalization scheme  in table~\ref{tab:pbendr}. 

\begin{table}[thb]
\begin{center}
\begin{tabular}{|c||c|c|c||c|c|c||c|c|c|}
\hline
& \multicolumn{3}{c||}{$\Lms^{(4)}=310\mev$} &
  \multicolumn{3}{c||}{$\Lms^{(4)}=340\mev$} &
  \multicolumn{3}{c| }{$\Lms^{(4)}=370\mev$} \\
\hline
$i$ & $r_i^{(0)}$ & $r_i^{(6)}$ & $r_i^{(8)}$ &
      $r_i^{(0)}$ & $r_i^{(6)}$ & $r_i^{(8)}$ &
      $r_i^{(0)}$ & $r_i^{(6)}$ & $r_i^{(8)}$ \\
\hline
0 &
   --3.167 &  14.781 &   1.815 &
   --3.192 &  15.974 &   1.686 &
   --3.215 &  17.277 &   1.546 \\
$X$ &
     0.546 &   0.026 &       0 &
     0.537 &   0.029 &       0 &
     0.529 &   0.032 &       0 \\
$Y$ &
     0.395 &   0.107 &       0 &
     0.385 &   0.113 &       0 &
     0.376 &   0.120 &       0 \\
$Z$ &
     0.626 &  --0.020 &  --12.574 &
     0.673 &  --0.021 &  --13.226 &
     0.723 &  --0.023 &  --13.927 \\
$E$ &
     0.189 &  --1.705 &   0.552 &
     0.179 &  --1.801 &   0.592 &
     0.169 &  --1.903 &   0.634 \\
\hline
\end{tabular}
\end{center}
\caption[]{Coefficients in the formula (\ref{eq:pbePi}) for various
$\Lms^{(4)}$ in the NDR scheme \cite{BJ03}.\label{tab:pbendr}}
\end{table}

\subsection{ Numerical Analysis}
The values of $\IM\lambda_t$ for various values 
of $1/R$ are given in table~\ref{BKL}. 
In table~\ref{tab:eps} we show the corresponding  results for $\epe$ 
as functions of the non-perturbative parameters in question. 
To this end we have set $m_s(m_c)= 100\mev$ which corresponds to
$m_s(2\gev)= 85\mev$, in the ballpark of some most recent lattice
calculations with dynamical fermions
\cite{Gupta}. The last year lattice values \cite{Wittig} and the 
QCD sum rules values \cite{Jamin} are slightly higher: 
$m_s(m_c)= 115\pm 20\mev$.

\begin{table}[htb]
\begin{center}
\begin{tabular}{|c||c|c||c|c||c|c|}
\hline
& \multicolumn{2}{c||}{$B_6^{(1/2)} = 1.00$} &
  \multicolumn{2}{c||}{$B_6^{(1/2)} = 1.15$} &
  \multicolumn{2}{c| }{$B_6^{(1/2)} = 1.30$} \\
\hline
$1/R $ & 
$B_8^{(3/2)}=0.8$ & $B_8^{(3/2)}=1.0$ &
$B_8^{(3/2)}=0.8$ & $B_8^{(3/2)}=1.0$ &
$B_8^{(3/2)}=0.8$ & $B_8^{(3/2)}=1.0$  \\
\hline
200 & 9.1 & 5.6  & 12.7 & 9.2 & 16.4 & 12.9 \\
250 & 11.0 & 7.7 & 14.8 & 11.5 & 18.6 & 15.3 \\
300 & 12.1 & 9.1 & 16.0 & 12.9 & 19.9 & 16.8 \\
400 & 13.4 & 10.6 & 17.3 & 14.5 & 21.3 & 18.5 \\
SM & 15.2 & 12.8 & 19.3 & 16.8 & 23.4 & 20.9 \\
\hline
\end{tabular}
\end{center}
\caption[]{The ratio $\epe$ in units of $10^{-4}$ for $\mt = 167 \gev$, 
$\Lms^{(4)}=340\mev$, 
$\ms(\mc)=100\mev$ and $m_d(\mc)=6\mev$ as a function of $1/R$ in $\gev$ and 
various values of $B_6^{(1/2)}$ and $B_8^{(3/2)}$.
\label{tab:eps}}
\end{table}

The results in table~\ref{tab:eps} are self explanatory. Dependently on 
the values of the parameters $B_6^{(1/2)}$ and $B_8^{(3/2)}$ one finds 
the value of $1/R$ that is favoured by the experimental data. Only for 
the case $(B_6^{(1/2)},B_8^{(3/2)})=(1.0,1.0)$   
the results for $\epe$ are outside
the experimental range in (\ref{eps}) independently of $1/R$ considered. As 
for $310\mev \le \Lms^{(4)} \le 370\mev$ the ratio $\epe$ is 
approximately proportional to $\Lms^{(4)}$, the results in 
table~\ref{tab:eps} can be at most changed by $\pm 10\%$. 

Clearly, the hadronic uncertainties in $(B_6^{(1/2)},B_8^{(3/2)})$ and 
$m_s(m_c)$ preclude any determination of $1/R$ from the data on $\epe$ 
at present. As generally $\epe$ is suppressed by the KK mode contributions 
relatively to the SM,  the ACD model is disfavoured 
by $\epe$ unless $(B_6^{(1/2)},B_8^{(3/2)})$ and $m_s(m_c)$ or generally 
the matrix elements of $Q_6$ and $Q_8$ are such that 
the SM estimates are above the data. This is the case for sufficiently 
low values of $m_s(m_c)$ and sufficiently high values of 
$B_6^{(1/2)}$. For these cases a suppression of $\epe$ by the KK 
modes found here is a welcome feature of the ACD model.

\section{Summary and Outlook}
In this paper we have calculated for the first time the contributions of 
the Kaluza-Klein (KK) modes to 
$B\to X_s~{\rm gluon}$, $B\to X_s\mu^+\mu^-$ and 
$K_L\to \pi^0e^+e^-$ and to the CP-violating ratio $\epe$
in the Appelquist, Cheng and Dobrescu (ACD) model with one universal extra 
dimension. We have also analyzed $B\to X_s\gamma$ that has been considered 
in the past \cite{AGDEWU}. While our result for this decay differs in details
from the latter paper, we confirm the suppression of the relevant branching 
ratio found by these authors.  
As a byproduct we have given a list of the required Feynman 
rules involving photons and gluons that have not been presented in the 
literature so far.
Moreover we have  generalized the background field method to 
five dimensions.  

 The nice property of this extension of the SM is the presence of
only a single new parameter, $1/R$.
This economy in new parameters should be 
contrasted with supersymmetric theories and models with an extended Higgs 
sector.
Taking  $1/R=300\GeV$ our findings are as follows:
\begin{itemize}
\item
The short distance function $Z$, relevant for $B\to X_s\mu^+\mu^-$, 
$K_L\to \pi^0e^+e^-$ and  $\epe$ is enhanced by $23\%$ relative to the 
SM value.
\item
The functions $D'$ and $E'$, relevant for $B\to X_s\gamma$ and 
$B\to X_s~{\rm gluon}$ are suppressed by $36\%$ and $66\%$, respectively.
The corresponding effects in the function $D$ are negligible and in $E$ 
phenomenologically irrelevant.
\item
$Br(B\to X_s\gamma)$ is suppressed by $20\%$. The phenomenological
implications of this result depend sensitively on the value of $m_c/m_b$ 
and on the experimental data, with the lower bound on $1/R$ being stronger 
for $m_c/m_b$=0.29 than for $m_c/m_b$=0.22. In fact, in the latter case the 
suppression of  $Br(B\to X_s\gamma)$ could be welcome and an upper bound on
$1/R$ could in principle be found when the experimental and theoretical 
uncertainties decrease.
\item
$Br(B\to X_s~{\rm gluon})$ is suppressed by $40\%$. The phenomenological 
relevance of this result is, in view of large hadronic uncertainties 
and the difficulty in extracting this branching ratio from the data, 
unclear 
at present.
\item
The perturbative part of $Br(B\to X_s\mu^+\mu^-)$ is enhanced by $12\%$.
This could be a welcome feature of the ACD model in view of the
data from Belle, that is above the SM expectations.
However, more interesting is the shift of the zero in the 
$A_{\rm FB}$ asymmetry from $\hat s_0=0.162$ to $\hat s_0=0.142$, 
that in view of small theoretical uncertainties could turn out to be 
a very important test of the ACD model.
\item
We have pointed out a correlation between
the zero $\hat s_0$ in  the $A_{\rm FB}$ asymmetry and 
$Br(B\to X_s\gamma)$ that should be valid in most models with minimal 
flavour violation.   
This correlation is shown in fig.~\ref{corrplot}.
\item
$Br(K_L\to \pi^0e^+e^-)$ is enhanced by less than $10\%$.
\item
$\epe$ is suppressed relative to the 
SM expectations with the size of the suppression depending 
sensitively on the hadronic matrix elements. Taking
$m_s(2\gev)=85\mev$ as suggested by the most recent lattice
calculations, we find that for several sets of
$(B_6^{(1/2)},B_8^{(3/2)})$ the ACD model can be made 
compatible with the experimental data, see table~\ref{tab:eps} for
details.  
\end{itemize}

These findings should be compared with the ones of \cite{BSW02}, where 
for $1/R=300\gev$ the following enhancements relative to the SM predictions 
due to enhanced $Z^0$-penguins have been found:
 $\kpn~(9\%)$, $\klpn~(10\%)$, 
$B\to X_{d}\nu\bar\nu~(12\%)$, $B\to X_{s}\nu\bar\nu~(21\%)$, 
$K_L\to\mu\bar\mu~(20\%)$, $B_{d}\to\mu\bar\mu~(23\%)$ and 
$B_{s}\to\mu\bar\mu~(33\%)$.

We observe that the impact of the KK modes on  
$B\to X_s\gamma$ is comparable to the one on  
$B\to X_{s}\nu\bar\nu$, 
$K_L\to\mu\bar\mu$ and $B_{d}\to\mu\bar\mu$. The corresponding impact
on $B\to X_s~{\rm gluon}$ is significantly larger and comparable to 
the one on $B_s\to \mu^+\mu^-$. On the whole, the main effects 
of the KK modes are felt in $Z^0$-penguins,
$\gamma$--magnetic penguins and chromomagnetic penguins. 

Moreover, the low compactification scale $200\gev$ 
that was fully compatible with the data on the decays considered there, 
is excluded by the $B\to X_s\gamma$  decay. For 
$1/R=200\gev$ the branching ratio $Br(B\to X_s\gamma)$ is suppressed 
relative to the SM by a factor of 1.6. This is clearly excluded. 
Of considerable 
interest is an even stronger suppression of $B\to X_s~{\rm gluon}$. 

Interestingly, the present data on $K^+\to\pi^+\nu\bar\nu$ and 
$B\to X_s\mu^+\mu^-$ are somewhat above the SM expectations and the
{\it enhancement} of the $Z^0$-penguins by the KK modes could be welcome 
in this context.
On the other hand, the present central experimental values for 
$Br(B\to X_s\gamma)$ are somewhat below 
the SM expectations and the {\it suppression} of the magnetic  
penguins by the KK modes could also be welcome.
However, in all these cases
the experimental and theoretical uncertainties have to be reduced in
order to be definitive about the need for these enhancements and suppressions.

Whether the suppression of  $B\to X_s~{\rm gluon}$ and of $\epe$ by the KK 
modes relative to the SM expectations is a welcome feature depends 
very strongly on hadronic uncertainties and in the case of
$B\to X_s~{\rm gluon}$ on the data.

Possibly, the most interesting result of our paper is the sizable downward 
shift of  the zero ($\hat s_0$) in the 
$A_{\rm FB}$ asymmetry. It should be emphasized that this shift has a
definite sign and the theoretical uncertainties in its value are small.

Our present analysis combined with our previous paper \cite{BSW02} 
completes the study 
of the most interesting FCNC processes in the ACD model. With only one 
additional parameter, the compactification scale, the pattern of various 
enhancements and suppressions relative to the SM expectations could be 
uniquely determined:
\begin{itemize}
\item
Enhancements: $K_L\to \pi^0e^+e^-$,  $\Delta M_s$,
 $\kpn$, $\klpn$, $B\to X_{d}\nu\bar\nu$, $B\to X_{s}\nu\bar\nu$, 
$K_L\to\mu^+\mu^-$, $B_{d}\to\mu^+\mu^-$, $B\to X_s\mu^+\mu^-$ and 
$B_{s}\to\mu^+\mu^-$.
\item
Suppressions: 
$B\to X_s\gamma$, $B\to X_s~{\rm gluon}$, the value of $\hat s_0$ in the 
forward-backward asymmetry and $\epe$.
\end{itemize}
Whether these enhancements and suppressions are required by the data or 
whether they exclude the ACD model with a low compactification scale, 
will depend 
on the precision of the forthcoming experiments and the efforts to decrease 
the theoretical uncertainties.

\section*{Acknowledgements}
We would like to thank first of all Christoph Greub but also 
Christoph Bobeth,  Paolo Gambino, 
 Uli Haisch, Gudrun Hiller, Matthias Jamin, Mikolaj Misiak and J\"org Urban for 
discussions  
related to their work within the SM on the decays considered in this paper  
and for providing the relevant NLO and NNLO computer programs that accelerated 
considerably our work. We also thank Gino Isidori and Frank Kr\"uger for 
informative 
discussions on the $B\to X_s\mu^+\mu^-$ decay.
This
research was partially supported by the German `Bundesministerium f\"ur 
Bildung und Forschung' under contract 05HT1WOA3 and by the 
`Deutsche Forschungsgemeinschaft' (DFG) under contract Bu.706/1-2.

\appendix
\pagebreak

\newsection{Background Field method in 5 dimensions}
\allowdisplaybreaks[1]
\label{backgroundfieldmethod}
In the calculation of the off-shell amplitude, we have used the method
of background fields. As in 1-PI diagrams the background fields only
appear on the external legs, there is no need to fix the gauge for the
background gauge fields.
With the appropriate choice of the gauge fixing for the quantum gauge
fields, we can ensure the invariance of the effective action with respect to
background field (BF) gauge transformations.
The general procedure in 4 dimensions is
described in \cite{Abbott}. Here we only state the SM case
and modify the gauge fixing to fit the ACD model.

The starting point is the action $S[\psi]$ before gauge fixing, where
$\psi$ denotes all gauge and matter fields in the action. We get the
BF action by the transformation
\be
  S[\psi]\longrightarrow S[\psi+\hat\psi],
\ee
where we have introduced the background fields $\hat\psi$. It
is modified by the gauge fixing to
\be\label{SBF}
  S_{\text{BF}}[\psi,\hat\psi] = S[\psi+\hat\psi] - \frac{1}{2\xi}\int
  d^dx~ \tilde\mcG\tilde\mcG + \text{ghost terms},
\ee
where $\tilde\mcG\tilde\mcG$ stands for all gauge fixing functionals.
As the terms with ghost fields are not relevant for us we will not
consider them here.

In the 4-dimensional electroweak SM, a convenient choice for the gauge
fixing functionals is
\be\label{GHiggsAbel}
  \tilde\mcG_{B,\text{SM}}[B,\phi,\hat\phi] = \partial_\mu B^\mu
  -ig'\xi\frac 12\left(\hat\phi^\dagger\phi -
    \phi^\dagger\hat\phi\right),
\ee
\be\label{GHiggs}
  \tilde \mcG^a_{A,\text{SM}}[A,\hat A,\phi,\hat\phi] = \partial_\mu
  A^{a\mu} + g_2 \epsilon^{abc}\hat A_\mu^b A^{c\mu}
  -ig_2\xi\frac 12\left(\hat\phi^\dagger\sigma^a\phi -
    \phi^\dagger\sigma^a\hat\phi\right),
\ee
with the Higgs fields
\begin{align}
  \hat\phi &= \frac {1}{\sqrt 2}\left(v+\hat\psi +
    i\hat\chi^a\sigma^a\right) \begin{pmatrix} 0\\1\end{pmatrix},
  &\phi &= \frac {1}{\sqrt 2}\left(\psi +
    i\chi^a\sigma^a\right) \begin{pmatrix} 0\\1\end{pmatrix},
\end{align}
where $\sigma^a$ are the common Pauli matrices. According to
(\ref{SBF}), the gauge fixing part of the Lagrangian is
\be\label{GFLag}
  \mcL_\text{GF} = -\frac{1}{2\xi}\tilde\mcG_B\tilde\mcG_B
  -\frac{1}{2\xi}\tilde\mcG_A^a\tilde\mcG_A^a.
\ee

With this specific gauge fixing, the BF action is invariant
under the BF gauge transformation\footnote{We omit the BF gauge
  transformation of the fermions which is just an ordinary gauge
  transformation.}
\begin{align}
  \label{bftB}
  \delta_\text{BF} \hat B_\mu &=
  \frac{1}{g'}\partial_\mu\beta,\\
  \label{bftA}
  \delta_\text{BF} \hat A_\mu^a &= f^{abc}\hat
  A_\mu^b\alpha^c + \frac{1}{g_2}\partial_\mu\alpha^a,\\
  \delta_\text{BF} \hat\phi &=
  i\left(\alpha^a\frac{\sigma^a}{2}+\frac{1}{2}\beta\right)\hat\phi,
\intertext{combined with a transformation of the quantum fields}
  \label{bftQ}
  \delta A_\mu^a &= f^{abc}A_\mu^b\alpha^c,\\
  \label{bftphi}
  \delta \phi &= i\left(\alpha^a\frac{\sigma^a}{2}+\frac{1}{2}\beta
  \right)\phi, 
\end{align}
where (\ref{bftQ}) and (\ref{bftphi}) are just unitary rotations in
the functional integral.

The analogous BF gauge transformation in 5 dimensions is
\begin{align}
  \delta_\text{BF} \hat B_M &= \frac{1}{\hat
    g'}\partial_M\beta,\\
  \delta_\text{BF} \hat A_M^a &= f^{abc}\hat
  A_M^b\alpha^c + \frac{1}{\hat
    g_2}\partial_M\alpha^a,\\
  \delta_\text{BF} \hat\phi &=
  i\left(\alpha^a\frac{\sigma^a}{2}+\frac{1}{2}\beta\right)\hat\phi
\intertext{combined with}
  \delta A_M^a &= f^{abc}A_M^b\alpha^c,\\
  \delta \phi &= i\left(\alpha^a\frac{\sigma^a}{2} + \frac{1}{2}\beta
  \right)\phi. 
\end{align}
Compared to 4 dimensions, the couplings $g'$ and $g_2$ have been
replaced by their 5-dimensional analogs $\hat g'$ and $\hat g_2$. All
fields and $\alpha^a$ and $\beta$ are now also functions of the
extra coordinate $y$ and the vector index $M$ can take the values
$M=0,1,2,3,5$.

In the ACD model, we have to add $-\xi\partial_5 B_5$ to
(\ref{GHiggsAbel}) and $-\xi\partial_5 A^a_5$ to (\ref{GHiggs}) in
order to diagonalize the mass matrices of the bosonic modes
\cite{BSW02}. However, this spoils BF gauge invariance, so we add
another term to (\ref{GHiggs}) to fix this and get
\be\label{GHiggsAbelACD}
  \tilde\mcG_{B,\text{ACD}}[B,\phi,\hat\phi] =
  \tilde\mcG_{B,\text{SM}}[B,\phi,\hat\phi]
  -\xi\partial_5 B_5,
\ee
\be\label{GHiggsACD}
  \tilde \mcG^a_{A,\text{ACD}}[A,\hat A,\phi,\hat\phi] =
  \tilde \mcG^a_{A,\text{SM}}[A,\hat A,\phi,\hat\phi]
  -\xi\partial_5 A^a_5-\xi\hat g_2 \epsilon^{abc} \hat A^b_5 A^c_5,
\ee
where it is understood that $g'$ and $g_2$ have been replaced by $\hat
g'$ and $\hat g_2$.

The use of (\ref{GHiggsAbelACD}) and (\ref{GHiggsACD}) ensures
invariance of all 1-PI diagrams under BF gauge transformations in 5
dimensions. However, as we are only interested in external zero-mode
fields, the last term in (\ref{GHiggsACD}) does not affect our
calculation.

In the diagrams for the calculation of the functions $E$ and $E'$, the
external gluon couples only to the quarks. As there are no fermions
involved in the gauge fixing, the background gluon couples to
quarks just like a quantum gluon, and there is no need to
specify a gauge fixing functional for QCD.
For completeness, we note that
\be\label{GHiggsQCD}
  \tilde \mcG^a_{G,\text{ACD}}[G,\hat G] = \partial_\mu
  G^{a\mu} + \hat g_s f^{abc}\hat G_\mu^b G^{c\mu}
  -\xi\partial_5 G^a_5-\xi\hat g_s f^{abc} \hat G^b_5 G^c_5
\ee
would be an appropriate choice.
Here $f^{abc}$ are the $SU(3)_C$ structure constants and $\hat g_s$ is
the strong coupling constant.

\newsection{Feynman Rules in the ACD Model: Photon and Gluons}
\label{Feynmanrules}
In this section we list the Feynman rules needed for the calculations
in this paper except for those already given in \cite{BSW02}. The
Feynman rules are derived in the 5d background field $R_\xi$-gauge
described in appendix~\ref{backgroundfieldmethod}. The rules
for vertices involving only quantum fields are the same as in the
conventional 5d $R_\xi$-gauge described in \cite{BSW02}. 

In order to simplify the notation, we omit the KK indices of the
fields. There is no 
ambiguity because in one-loop calculations at least one field is
always a zero-mode. In the vertex rules given below, this is the
photon $A$, the gluon $G$ and their background equivalents.
Due to KK parity conservation, the other two
fields have equal KK mode number, i.e. either zero or $n \geq 1$.

Fermion zero-modes have substantially different Feynman rules than
their KK excitations.
The up-type quarks $\mcQ_i$ and $\mcU_i$
are always supposed to be $(n\geq 1)$-modes, while the
zero-mode is labeled $u$.
The generation index $i$ can take the values $i=u,c,t$.

In the vertices
below, $S^\pm$ stands for the scalar modes $G^\pm$ and $a^\pm$.
All momenta and fields are assumed to be incoming.
The Feynman rules for the vertices are:

\vspace{.5cm}
\begin{minipage}{38mm}
  \fbox{\includegraphics[]{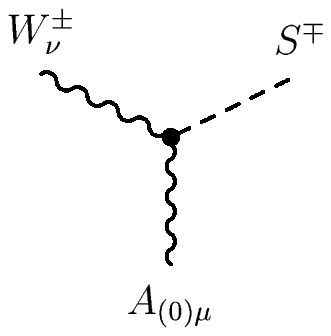}}
\end{minipage}
\begin{minipage}{12cm}
  $\displaystyle = g_2 \sw M\kkx Wn g_{\mu\nu} C.$
\end{minipage}

\begin{alignat}{4}
  &A W^+ G^-  &\dpkt C &= 1,
  &\hspace{24ex}&A W^- G^+  &\dpkt C &= -1,\\
  &A W^+ a^-  &\dpkt C &= 0,
  &&A W^- a^+  &\dpkt C &= 0,\\[3ex]
  &\hat A W^+ G^-  &\dpkt C &= 0,
  &&\hat A W^- G^+  &\dpkt C &= 0,\\
  &\hat A W^+ a^-  &\dpkt C &= 0,
  &&\hat A W^- a^+  &\dpkt C &= 0.
\end{alignat}

\vspace{.5cm} 
\begin{minipage}{42mm}
  \fbox{\includegraphics[]{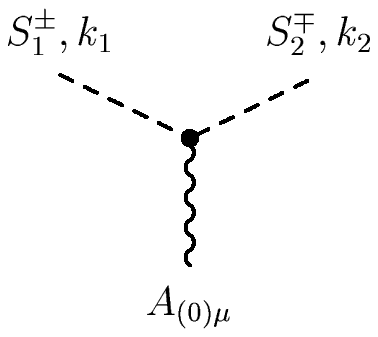}}
\end{minipage}
\begin{minipage}{12cm}
  $\displaystyle = -i g_2 \sw (k_2-k_1)_\mu C.$
\end{minipage}

\begin{alignat}{4}
  &A G^+G^-   &\dpkt C &= 1,
  &\hspace{24ex}&A a^+a^-  &\dpkt C &= 1,\\
  &A G^+a^-   &\dpkt C &= 0,
  &&A a^+G^-  &\dpkt C &= 0,
\end{alignat}
and the same values of $C$ for the analogous vertices with a background
photon $\hat A$. 

\vspace{.5cm}
\begin{minipage}{44mm}
  \fbox{\includegraphics[]{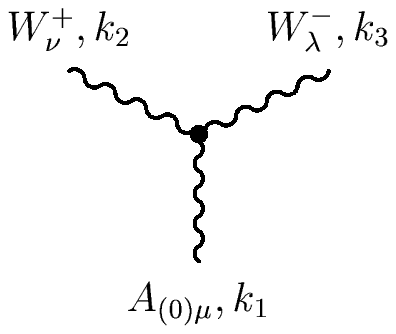}}
\end{minipage}
\begin{minipage}{12cm}
  $\displaystyle = ig_2\sw C.$
\end{minipage}

\begin{alignat}{2}
  &A W^+W^-  &\dpkt C &= g_{\mu\nu} (k_2-k_1)_\lambda + g_{\mu\lambda}
  (k_1-k_3)_\nu + g_{\lambda\nu} (k_3-k_2)_\mu,\\
  &\hat A W^+W^-  &\dpkt C &= g_{\mu\nu} (k_2
      -k_1+\frac{1}{\xi}k_3)_\lambda + g_{\mu\lambda} (k_1 -k_3 -
      \frac{1}{\xi}k_2)_\nu\\
  &&& + g_{\lambda\nu} (k_3 -k_2)_\mu.
\end{alignat}

\vspace{.5cm}
\begin{minipage}{36mm}
  \fbox{\includegraphics[]{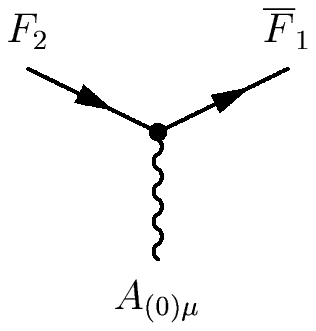}}
\end{minipage}
\begin{minipage}{12cm}
  $\displaystyle  = i g_2 \sw \gamma_\mu C.$
\end{minipage}

\begin{alignat}{4}
  & A \overline u_i u_i  &\dpkt &C = \frac{2}{3},
  &\hspace{24ex}&&&\\
  & A \overline \mcQ_i \mcQ_i     &\dpkt 
  &C = \frac{2}{3},
  && A \overline \mcU_i \mcU_i     &\dpkt 
  &C = \frac{2}{3},\\
  & A \overline \mcQ_i \mcU_i     &\dpkt 
  &C = 0,
  && A \overline \mcU_i \mcQ_i     &\dpkt 
  &C = 0,
\end{alignat}
and the same values of $C$ for the analogous vertices with a background
photon $\hat A$. 

\vspace{.5cm}
\begin{minipage}{38mm}
  \fbox{\includegraphics[]{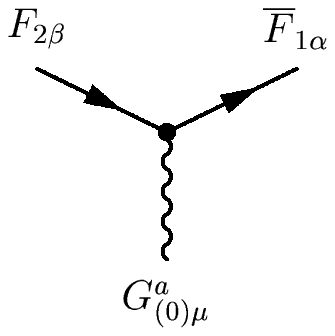}}
\end{minipage}
\begin{minipage}{12cm}
  $\displaystyle  = i g_s T^a_{\alpha\beta} \gamma_\mu C.$
\end{minipage}

\begin{alignat}{4}
  & G \overline u_i u_i  &\dpkt &C = 1,
  &\hspace{24ex}&&&\\
  & G \overline \mcQ_i \mcQ_i     &\dpkt 
  &C = 1,
  && G \overline \mcU_i \mcU_i     &\dpkt 
  &C = 1,\\
  & G \overline \mcQ_i \mcU_i     &\dpkt 
  &C = 0,
  && G \overline \mcU_i \mcQ_i     &\dpkt 
  &C = 0,
\end{alignat}
and the same values of $C$ for the analogous vertices with a background
gluon $\hat G$.


\newpage

\newcommand{\np}[3]{Nucl.~Phys. {\bf B#1} (#2) #3}
\newcommand{\pl}[3]{Phys.~Lett. {\bf B#1} (#2) #3}
\newcommand{\pr}[3]{Phys.~Rev.  {\bf D#1} (#2) #3}
\newcommand{\prl}[3]{Phys.~Rev. Lett. {\bf #1} (#2) #3}
\newcommand{\prp}[3]{Phys.~Rept. {\bf #1} (#2) #3}
\newcommand{\zpc}[3]{Z.~Phys. {\bf C#1} (#2) #3}
\newcommand{\hep}[2]{[arXiv:hep-#1/#2]}

\renewcommand{\baselinestretch}{0.95}

\vfill\eject

\end{document}